\documentclass[lettersize,journal]{IEEEtran}
\usepackage{amsmath,amsfonts}
\usepackage[ruled, linesnumbered]{algorithm2e}
\usepackage{array}
\usepackage[caption=false,font=normalsize,labelfont=sf,textfont=sf]{subfig}
\usepackage{textcomp}
\usepackage{stfloats}
\usepackage{url}
\usepackage{verbatim}
\usepackage{graphicx}
\usepackage{cite}
\usepackage{multirow}
\usepackage{amssymb}
\usepackage{amsmath}
\usepackage{threeparttable}

\usepackage{tikz,xcolor,hyperref}% Make Orcid icon
\hypersetup{
  colorlinks=true}
\definecolor{lime}{HTML}{A6CE39}
\DeclareRobustCommand{\orcidicon}{%
    \begin{tikzpicture}
    \draw[lime, fill=lime] (0,0) 
    circle [radius=0.16] 
    node[white] {{\fontfamily{qag}\selectfont \tiny ID}};    \draw[white, fill=white] (-0.0625,0.095) 
    circle [radius=0.007];    \end{tikzpicture}
    \hspace{-2mm}}
\foreach \x in {A, ..., Z}{%
    \expandafter\xdef\csname orcid\x\endcsname{\noexpand\href{https://orcid.org/\csname orcidauthor\x\endcsname}{\noexpand\orcidicon}}
    }

\begin{document}

\title{DNFS-VNE: Deep Neuro Fuzzy System Driven Virtual Network Embedding}
% \title{Virtual Network Embedding Meets Deep Fuzzy Neural Systems: An Initial Exploration of Interpretable Resource Allocation}
% \vspace{-2cm}
\author{Ailing Xiao\orcidA{},~\IEEEmembership{Member,~IEEE}, Ning Chen\orcidD{},~\IEEEmembership{Student Member,~IEEE},
Sheng~Wu\orcidH{},~\IEEEmembership{Member,~IEEE}, \\Peiying~Zhang\orcidB{},~\IEEEmembership{Member,~IEEE}, Linling Kuang\orcidE{},~\IEEEmembership{Member,~IEEE}, Chunxiao Jiang\orcidC{},~\IEEEmembership{Fellow,~IEEE}
% Wei Zhang\orcidI{},~\IEEEmembership{Fellow,~IEEE}

\thanks{
   This work is partially supported by the National Natural Science Foundation of China under Grant 62341104, 62325108, and 62341131; partially supported by BUPT Excellent Ph.D. Students Foundation under Grant CX20241074; partially supported by the R\&D Program of Beijing Municipal Education Commission under Grant KM202110009004; partially supported by Start-up Fund for Newly Introduced Teacher under Grant BUPT2024RC01; partially supported by the Natural Science Foundation of Shandong Province under Grant ZR2023LZH017 and ZR2022LZH015.
    (\textit{Corresponding author:} \textit{Sheng Wu}.)
    }
\thanks{Ailing Xiao, Ning Chen, and Sheng Wu are with School of Information and Communication Engineering, Beijing University of Posts and Telecommunications, Beijing 100876, China (\textit{e-mails: xiao\_ailing@bupt.edu.cn, nchen@bupt.edu.cn, thuraya@bupt.edu.cn}).}

\thanks{Peiying Zhang is with Qingdao Institute of Software, College of Computer Science and Technology, China University of Petroleum (East China), Qingdao 266580, China (\textit{e-mail: zhangpeiying@upc.edu.cn}).}
% \thanks{Suzhi Cao is with Key Laboratory
% of Space Utilization, Technology and Engineering Center for Space
% Utilization, Chinese Academy of Sciences, Beijing 100094, China (\textit{e-mail: caosuzhi@csu.ac.cn}).}

\thanks{
Linling Kuang and Chunxiao Jiang are with Tsinghua Space Center, Tsinghua University, and
also with the Beijing National Research Center for Information Science
and Technology, Tsinghua University, Beijing 100084, China (\textit{e-mails:
kll@tsinghua.edu.cn, jchx@tsinghua.edu.cn}).
}
% \thanks{Wei Zhang is with the School of Electrical Engineering and Telecommunications, University of New South Wales, Sydney, NSW 2052, Australia (\textit{e-mail: w.zhang@unsw.edu.au}).}
}

% \markboth{IEEE Internet of Things Journal}%


\maketitle

\begin{abstract}
By decoupling substrate resources, network virtualization (NV) is a promising solution for meeting diverse demands and ensuring differentiated quality of service (QoS). In particular, virtual network embedding (VNE) is a critical enabling technology that enhances the flexibility and scalability of network deployment by addressing the coupling of Internet processes and services. However, in the existing deep neural networks (DNNs)-based works, the black-box nature DNNs limits the analysis, development, and improvement of systems. For example, in the industrial Internet of Things (IIoT), there is a conflict between decision interpretability and the opacity of DNN-based methods. In recent times, interpretable deep learning (DL) represented by deep neuro fuzzy systems (DNFS) combined with fuzzy inference has shown promising interpretability to further exploit the hidden value in the data. Motivated by this, we propose a DNFS-based VNE algorithm that aims to provide an interpretable NV scheme. Specifically, data-driven convolutional neural networks (CNNs) are used as fuzzy implication operators to compute the embedding probabilities of candidate substrate nodes through entailment operations. And, the identified fuzzy rule patterns are cached into the weights by forward computation and gradient back-propagation (BP). Moreover, the fuzzy rule base is constructed based on Mamdani-type linguistic rules using linguistic labels. In addition, the DNFS-driven five-block structure-based policy network serves as the agent for deep reinforcement learning (DRL), which optimizes VNE decision-making through interaction with the environment. Finally, the effectiveness of evaluation indicators and fuzzy rules is verified by simulation experiments.
\end{abstract}


% \vspace{-0.3cm}
\section{Introduction}
\subsection{Background and Motivation}
\IEEEPARstart{N}{etwork} virtualization (NV) is the primary solution to Internet rigidity through the slicing and decoupling of network functions, applications, and services~\cite{10012414, 10413579, 9717289}. Through virtual network embedding (VNE) technology, infrastructure providers (InPs) can supply multiple sets of network services (which are defined as virtual network requests, VNRs) for multiple sets of Internet service providers (ISPs) to satisfy the needs of the network in terms of plurality, diversification, personalization, and high quality of service (QoS)~\cite{yan2020automatic}. 

Deep learning (DL) techniques, represented by deep neural networks (DNNs), have been maturely applied to various fields and successfully solved various industrial problems~\cite{10077713,9520818}. Up to now, the existing VNE algorithms are all based on artificial intelligence (AI) community methods, and all of them have also achieved good results~\cite{10263775}. However, due to their black-box nature, DNNs lack an interpretable exploration of data association, which limits the analysis and development of systems~\cite{10162187}. For example, in the Industrial Internet of Things (IIoT), although there are some advanced resource allocation methods~\cite{10413579}, the need for interpretability of decisions is becoming increasingly prominent, which not only helps to improve the transparency and interpretability of the system but also enhances researchers' trust in decisions, thereby optimizing resource utilization efficiency and reducing potential risks. The fuzzy system is a system that defines input, output, and state variables on fuzzy sets. It can mimic human comprehensive inference to deal with fuzzy information processing problems that are difficult to solve by conventional mathematical methods~\cite{7938307}. A comparison of the two is shown in Table~\ref{tab1}, and it can be found that each has advantages and disadvantages in the expression, storage, application, and acquisition of knowledge. Therefore, fuzzy neural networks (FNNs) combine fuzzy systems with DNNs, fully taking into account the complementary nature of the two, and show excellent results in dealing with large-scale fuzzy application problems~\cite{9669061}. This class of systems based on the hybrid approach of FNNs is called deep neuro fuzzy systems (DNFS), which combines logical reasoning, linguistic computation, and nonlinear dynamics, and has the capabilities of learning, association, recognition, adaptive and fuzzy information processing~\cite{8352739}, etc. To elaborate the potential associations to further explore the hidden value in the data, DNFS is widely used in various scientific researches such as communications, transportation, healthcare, etc., and exhibits higher interpretability and better decision-making~\cite{talpur2023deep}.

\begin{table*}[]
\centering
\caption{Comparison of deep neural networks and fuzzy system.}
\label{tab1}
\renewcommand\arraystretch{1.2}
\resizebox{\textwidth}{15mm}{\begin{tabular}{|l|l|l|}
\hline
\textbf{Comparison}      & \textbf{Neural Networks}                                           & \textbf{Fuzzy System}                                            \\ \hline
Basic Composition        & Multiple neurons                                                   & Fuzzy rule                                                       \\ 
Knowledge Acquisition    & Samples, algorithm examples                                        & Expert knowledge, logical reasoning                              \\ 
Knowledge Representation & Distributed representation                                         & Membership function                                              \\ 
Reasoning Mechanism      & Learning function self-control, parallel computing, fast speed     & Combination of fuzzy rules, heuristic search, slow speed         \\ 
Reasoning Operation      & Superposition of neurons                                           & Operation of membership function                                 \\ 
Adaptability             & Learning by adjusting weights, high fault tolerance                & Inductive learning, low fault tolerance                          \\ 
Advantage                & Self-organization, high fault tolerance and generalization ability & Can use expert experience, easy to understand, less calculation  \\ 
Disadvantages            & Black-box model, difficult to understand, heavy calculation        & Difficult to learn, increased fuzziness in the reasoning process \\ \hline
\end{tabular}
}
\end{table*}
Based on the above motivations, a DNFS-based VNE algorithm (DNFS-VNE) is proposed in this work to enhance the interpretability of the technique, clarify hidden patterns and associations, and facilitate system development. To our knowledge, this is one of the first explorations of the interpretability of the VNE algorithm using DNFS.
\vspace{-0.2cm}
\subsection{Overview of Related Work}
The VNE algorithms in the current researches mainly consist of two categories, namely heuristic strategies and AI-based strategies~\cite{drone2023}. Among them, heuristic strategies employ numerical optimization algorithms such as mixed integer programming (MIP), dynamic programming, and so on~\cite{10413579}. AI-based strategies employ reinforcement learning (RL), deep reinforcement learning (DRL), and other AI algorithms with powerful nonlinear fitting capabilities to better model the mathematical problem, dynamic characterization, and decision optimization of complex physical environments~\cite{9766416}.
\subsubsection{Heuristic Strategies}
The most classical algorithm to rank the substrate nodes based on their network topology and resource attributes applying the Markov Random Walk model was proposed by Cheng \textit{et al.}~\cite{cheng2011virtual}. And, they presented such node topology-aware VNE algorithm: NodeRank-VNE. In addition, they used the shortest path algorithm and breadth-first search (BFS) for virtual network embedding, respectively.
Another classic algorithm is the MIP-based VNE algorithm proposed by Chowdhury \textit{et al.}~\cite{5061987}. In this algorithm, the VNE problem is formulated as MIP, and two VNE algorithms, D-ViNE and R-ViNE, are designed by using deterministic and randomized rounding by relaxing the integer constraints. It is worth noting that these studies aim to maximize acceptance rate and revenue of the VNE algorithm.
Moreover, it is proved that VNE is a typical NP-hard optimization problem.
% Genetic algorithm (GA) is an optimization algorithm that mimics natural evolution and is used to solve complex optimization problems by searching for optimal solution space. Based on this starting point, our previous work~\cite{zhang2019virtual} proposed a VNE algorithm (VNE-MGA) based on a modified genetic algorithm and demonstrates improvements in terms of VNR acceptance rate and long-term average revenue.
Since previous works lacked consideration of storage resource constraints, and efficient utilization of storage resources can alleviate bandwidth consumption, one of our previous works~\cite{7976281} firstly proposed a VNE strategy based on 3D resource constraints of network, computing, and storage. Specifically, two heuristic VNE algorithms are designed based on the node mapping strategy: NRM-VNE and RCR-VNE.

Although heuristic strategies provided an effective solution for the early VNEs, such strategies have been difficult to apply to today's complex and variable substrate networks. Moreover, these static VNE algorithms assume that the VNR is known, which is impractical for time-varying service requests in networks. Moreover, the artificially customized embedding rules used limit the effectiveness of the strategy.
\subsubsection{AI-based Strategies}
For the problem that traditional VNE algorithms following static mechanisms are difficult to adapt to dynamic substrate network environments and some RL-assisted VNE algorithms ignore the continuity of node embedding, Yao \textit{et al.}~\cite{yao2020continuous} proposed an RL-based continuous decision-making VNE algorithm (CDRL). Specifically, this work converts the continuous node embedding process into a recurrent neural network (RNN)-based complex time-series problem and updates the parameters using a policy gradient algorithm. The final results demonstrate the superiority of CDRL in terms of evaluation metrics.
% Due to the dynamics of the topology and resources of the physical network as well as the VNR pose challenges to the practicality of the VNE algorithm, Zhang \textit{et al.}~\cite{9475485} applied graph neural networks (GNN) to the VNE algorithm and proposed a VNE algorithm (GCN-RL-VNE) combining RL with graph convolutional networks (GCN). Specifically, to improve the dynamics of the algorithm, this work formulated a novel fitness matrix and objective function. The good flexibility of GCN-RL-VNE is demonstrated by changing the substrate network environment and VNR resource properties.

In previous works, RL and DL provide better decision-making and perception for VNE algorithms, respectively. In addition, the emerging DRL combines the advantages of both by extracting environmental features through the nonlinear fitting ability of DNNs and guiding algorithm optimization through the interaction of the agent with the environment and incorporating feedback from reward mechanisms~\cite{zhang2022blockchain}. Therefore, the DRL paradigm has been the mainstream means of VNE algorithms in recent years~\cite{8845171}. Aiming at the problem that the algorithm needs to detect complex dynamic environments and provide automatic embedding solutions time-varyingly during runtime, Yan \textit{et al.}~\cite{yan2020automatic} proposed an automatic VNE algorithm combining GCN and DRL. The algorithm designs a multi-objective function and a parallel DRL architecture, and the results show that it has superior performance and robustness. To improve the resource utilization of vehicular fog computing networks, one of our previous works proposed a DRL-based VNE algorithm incorporating spectral graph theory~\cite{CHEN2022109931}. Specifically, a four-layer policy network based on spectral graph convolution was designed to compute the embedding probability of the substrate nodes, and optimization was guided by formulated reward signals.
In addition, to address the real-time, dynamic, and privacy issues of the substrate network, a VNE algorithm based on horizontal federated learning was proposed for the first time in our previous work~\cite{10132867}. Specifically, the server and the DRL model are respectively deployed in the global domain and the local domain. Among them, the global is responsible for aggregating and sharing parameters, and the local is responsible for focusing on local resource optimization, which significantly improves the algorithm efficiency while ensuring~privacy.

Although AI-based VNE algorithms are a big step forward from heuristic algorithms in terms of algorithmic dynamics and performance, the black-box nature of these AI-strategy restricts the exploration of the interpretability of potential data associations of the VNE algorithms, which limits the improvement and development of the related systems. As a result, the VNE community awaits an attempt at interpretable~algorithms.

\vspace{-0.2cm}
\subsection{Contribution}
To summarise, the innovations and contributions of this work are as follows:

1. Targeting research on multi-domain substrate networks, based on the DNFS paradigm, we propose the DNFS-VNE structure with five blocks. Specifically, we employ a DL model based on convolutional neural networks (CNNs) as a data-driven fuzzy implication operator for entailment operations. Further, based on the policy gradient algorithm, the DNFS-driven five-block structure acts as the agent for DRL to optimize decision-making by interacting with the~environment.

2. DNFS-VNE outputs inferred membership values, which are then aggregated and defuzzified to derive the embedding probabilities of substrate nodes. Furthermore, the identified fuzzy rule patterns are cached into weights by forward computation and gradient back-propagation. And, the fuzzy rule base is constructed by Mamdani-type linguistic rules between the antecedent layer and the consequent layer of the CNN-based fuzzy rule implication branch, using a set of linguistic labels to elucidate fuzzy implication principles.

3. Experiment verification demonstrates that DNFS-VNE outperforms other algorithms in terms of long-term average revenue, long-term average revenue-cost ratio, and VNR acceptance success rate, all of which are widely employed in VNE. More importantly, DNFS-VNE provides an interpretable solution to the previous black-box model of the VNE.

The content of this work is organized as follows: in Section~\ref{sec2}, the problem definition, problem modeling, DNFS paradigm, and related indicators have been presented; in Section~\ref{sec3}, the model details have been introduced, including model composition and learning process; in Section~\ref{sec4}, we have performed simulation experimental verification and analysis; Finally, in Section~\ref{sec5}, we have summarised and looked forward this work.
% \vspace{-0.3cm}
\section{Problem Definition and Modelling}\label{sec2}
As mentioned earlier, the main goal of VNE is to efficiently decouple substrate network resources for allocation to different VNRs. In terms of quantitative indicators, the main purpose of VNE is to improve the algorithm acceptance rate and revenue and reduce cost.
% \vspace{-0.3cm}
\subsection{Problem Definition}

\begin{figure}
    \centering
    \includegraphics[width=0.7\linewidth]{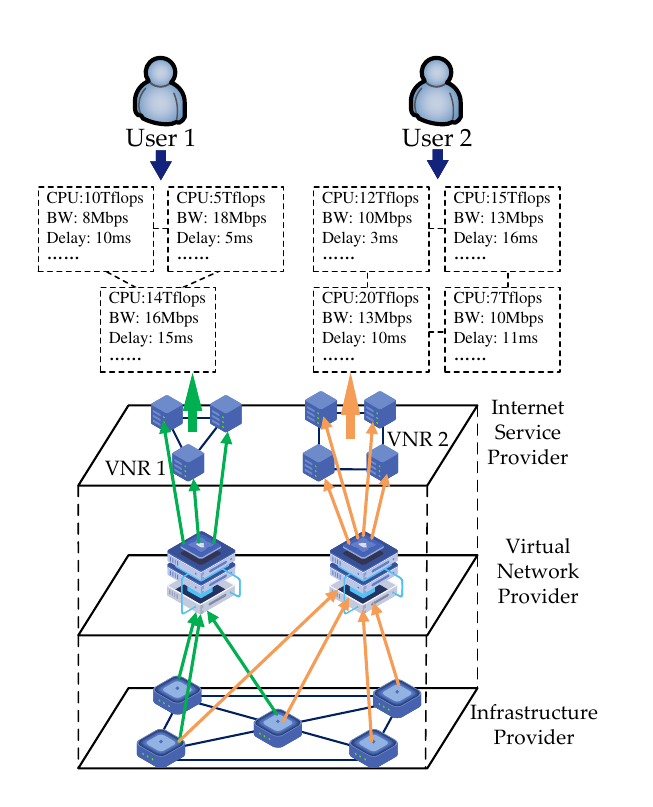}
    \caption{Schematic diagram of the VNE process that the VNRs of two users are embedded into the substrate network, where the numbers indicate the relevant network metrics.}
    \label{fig:1}
\end{figure}
A schematic diagram of the VNE process in which two users' VNRs are embedded into the substrate network is shown in Fig.~\ref{fig:1}.
From the figure, it can be found that for the ISP, when a user makes a service request, it creates VNRs that satisfy the corresponding demand (\textit{e.g.}, the numbers in the figure). For the InP, under the premise of the limited substrate network resources, it should respond to as many requests as possible and as reasonably as possible when a stable network flow sends service requests. Therefore, the VNE problem is defined as a problem of how to allocate the limited substrate network resources to VNRs more efficiently and reasonably.
In addition, the VNE problem is obviously an NP-hard problem~\cite{fischer2013virtual, CHEN2022109931}. Therefore, many researchers are constantly trying novel solutions, exploring and figuring out.
% \vspace{-0.3cm}
\subsection{Problem Modelling}
\begin{table}[h]
\vspace{-0.3cm}
\caption{Related notations definition.}
\label{tab:2}
\centering
\renewcommand\arraystretch{1.3}
\begin{tabular}{|l|l|l|}
\hline
\textbf{Network}                             & \textbf{Notation} & \textbf{Definition} \\ \hline
\multirow{5}{*}{Substrate Network $\mathsf{S}$}  
& $\boldsymbol{n}$ & Substrate nodes\\
& $\boldsymbol{l}_{\mathsf{intra}}$ & Intra-domain substrate links\\
& $\boldsymbol{l}_{\mathsf{inter}}$ & Inter-domain substrate links\\
& $\boldsymbol{r}_{\mathsf{n}}$ & Resource of substrate nodes\\
& $\boldsymbol{r}_{\mathsf{l}}$ & Resource of substrate links\\
\hline
\multirow{4}{*}{Virtual Network Request $\mathsf{V}$} 
& $\boldsymbol{n}_{\mathsf{v}}$  &    Virtual nodes        \\
& $\boldsymbol{l}_{\mathsf{v}}$ & Virtual links\\
& $\boldsymbol{r}_{\mathsf{n,v}}$ & Resource of virtual nodes\\
& $\boldsymbol{r}_{\mathsf{l,v}}$ & Resource of virtual links \\
 \hline
\end{tabular}

\end{table}
Targeting research on widely used multi-domain substrate networks, to study the VNE problem, it is necessary to model the network to clarify the process of the problem. Specifically, the related notations definition used in this work is displayed in Table~\ref{tab:2}. Among them, both the substrate network and the $i$-th VNR are modeled as weighted undirected graphs, as follows,
\begin{align}
    \mathcal{S} &= \{\boldsymbol{n}, \boldsymbol{l}_{\mathsf{intra}}, \boldsymbol{l}_{\mathsf{inter}}, \boldsymbol{r}_{\mathsf{n}}, \boldsymbol{r}_{\mathsf{l}}\},\\
    \mathcal{V}(i) &= \{\boldsymbol{n}_{\mathsf{v}}(i), \boldsymbol{l}_{\mathsf{v}}(i), \boldsymbol{r}_{\mathsf{n,v}}(i), \boldsymbol{r}_{\mathsf{l,v}}(i)\}.
\end{align}

Furthermore, similar to previous works, we use computing (denoted as $c$) resources as an example to represent node resource attributes and bandwidth (denoted as $w$) resources as an example to represent link resource attributes.

Therefore, as analyzed above, the VNE problem can be modeled abstractly as follows,
\begin{equation}
\begin{aligned}
    \text{for}\ i=1,2,\cdots,|\mathsf{V}|:
    \mathcal{S}' \rightarrow \mathcal{V}(i) (\text{from}\ t_i^{\mathsf{s}}\  \text{to}\ t_i^{\mathsf{e}}),
\end{aligned}
\end{equation}
where $|\mathsf{V}|$ indicates the number of VNRs, $\mathcal{S}'$ indicates the subset of the substrate network, $\mathcal{V}(i)$ indicates the $i$-th VNR, and $t_i^{\mathsf{s}}$ and $t_i^{\mathsf{e}}$ indicate the start time and end time of the $\mathcal{V}(i)$ life cycle, respectively.
To sum up, in other words, the VNE process is to effectively allocate part of the substrate network equipment and resources to each VNR within its life cycle to meet the corresponding network service requirements.

\begin{figure}
    \centering
    \includegraphics[width= 0.9\linewidth]{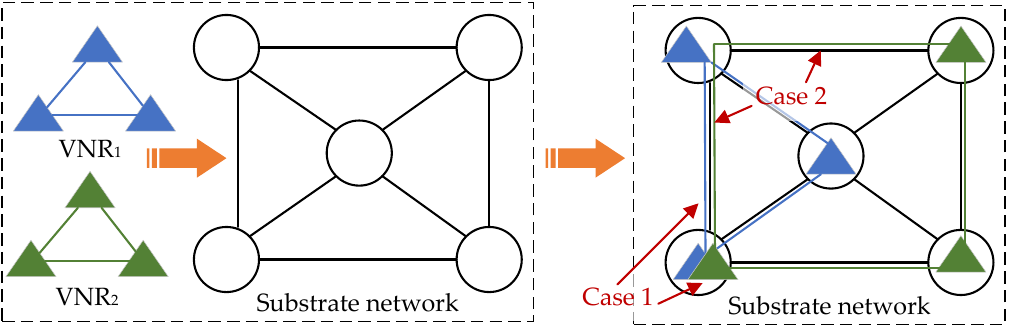}
    \caption{Schematic diagram of the corresponding relationship between substrate resources and virtual resources.}
    \label{fig:corr}
\end{figure}
It should be noted that in multi-layer nested virtualization scenarios since the lowest-level network resources are physical, the substrate network resources may also be virtual. However, its resource allocation can be easily derived from the general scenario, so these cases are not considered in this work. In addition, the substrate network resource can be hosted in multiple virtual networks as long as sufficient resources can be allocated and related constraints are met. It can be found that the correspondence between substrate nodes and virtual nodes is $1:n$, as shown in Case $1$ in Fig.~\ref{fig:corr}. Moreover, a virtual link mapping may span multiple substrate links, so the correspondence between substrate links and virtual links is $n:m$, as shown in Case $2$ in Fig.~\ref{fig:corr}.
% \vspace{-0.3cm}
\subsection{DNFS Paradigm}\label{sec2c}
\begin{figure}[!htbp]
    \centering
    \includegraphics[width=\linewidth]{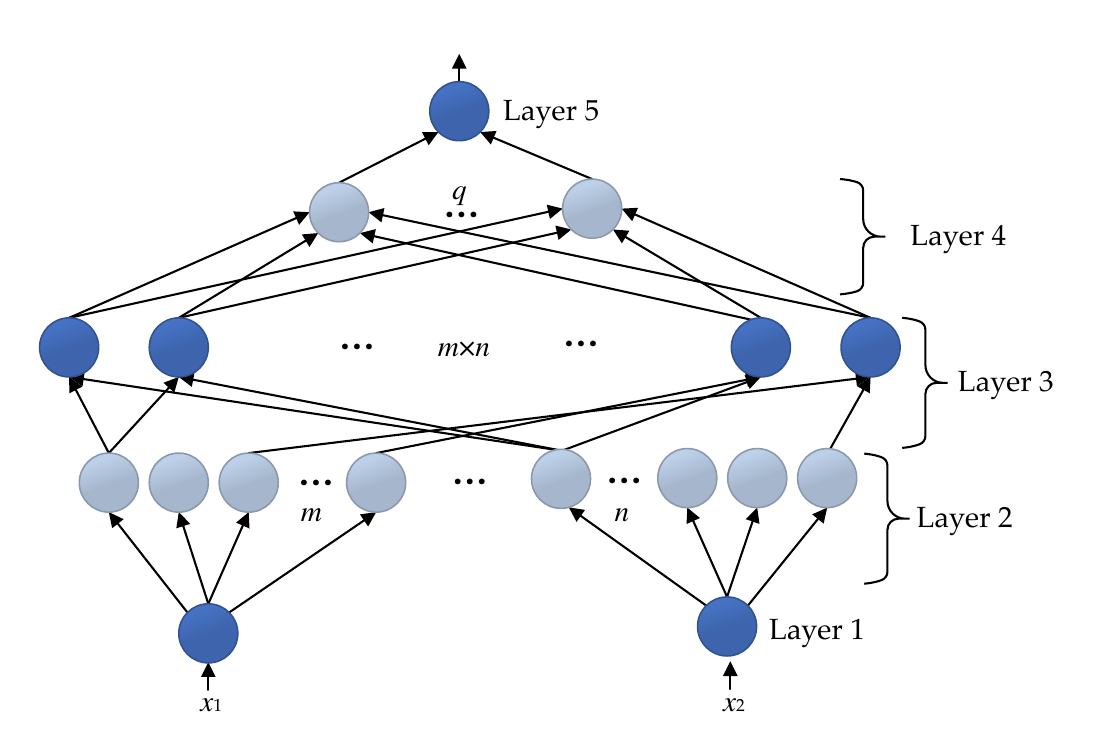}
    \caption{The five-layer feed-forward network structure of DNFS.}
    \label{fig:2}
\end{figure}
The most studied and most applied solution in the existing work is to customize DNFS as a five-layer feed-forward network structure. The specific structure is shown in Fig.~\ref{fig:2}, where each layer is,
\begin{itemize}
    \item Layer $1$: Input layer. The number of nodes is the number of input variables;
    \item Layer $2$: Fuzzification layer. It is the membership function layer of the input variables to realize the fuzzy of the input variables. For example, the size of fuzzy sets for $x_1$ and $x_2$ are $m$ and $n$, respectively;
    \item Layer $3$: Fuzzy rule layer. Each node in this layer is only connected to a unique node in each fuzzy set. The number of nodes in this layer is the number of fuzzy rules. For example, there are $m\times n$ nodes;
    \item Layer $4$: Defuzzification layer. This layer is fully connected and the number of nodes is the number $q$ of fuzzy divisions of the output variable;
\item Layer $5$: Output layer, also known as the clarity layer. This layer is a fully connected layer whose number of nodes is the number of output variables.
\end{itemize}
Therefore, DNFS essentially uses the learning method of DNNs to automatically design and adjust the design parameters of the fuzzy system according to the input and output learning samples, to realize the self-learning and self-adaptive functions of the fuzzy system.
% \vspace{-0.5cm}
\subsection{General Evaluation Indicators}
As mentioned earlier, improving the algorithm acceptance rate and revenue and reducing cost are the optimization goals of the VNE algorithm~\cite{fischer2013virtual,10040224}. That is, more resource allocation revenue is obtained with less substrate network resource cost. Therefore, the following equations are used to intuitively measure cost and revenue for $\mathcal{V}(i)$,
\begin{align}
&\ \begin{aligned}
    \zeta_i(t_i^{\mathsf{s}},t_i^{\mathsf{e}})
    &=\sum_{j=1}^{j=| \boldsymbol{n}_{\mathsf{v}}(i)|} {r}_{\mathsf{n},\mathsf{v}}(i,j) +\sum_{k=1}^{ k=|\boldsymbol{l}_{\mathsf{v}}(i)|} h\times {r}_{\mathsf{l},\mathsf{v}}(i,k),\\
    &=\sum_{j=1}^{j=| \boldsymbol{n}_{\mathsf{v}}(i)|} {c}_{\mathsf{n},\mathsf{v}}(i,j) +\sum_{k=1}^{ k=|\boldsymbol{l}_{\mathsf{v}}(i)|} h\times {w}_{\mathsf{l},\mathsf{v}}(i,k),\label{eq3}
    % &=\sum_{n^{\text{R}_i}\in \mathbf{N}^{\text{R}_i}}\mathbb{C}^{\text{R}_i}(n^{\text{R}_i})+\sum_{l^{\text{R}_i}\in \mathbf{L}^{\text{R}_i}} hops\times \mathbb{B}^{\text{R}_i}(l^{\text{R}_i})
\end{aligned}\\
&\ \begin{aligned}
    \xi_i(t_i^{\mathsf{s}},t_i^{\mathsf{e}})
    &=\sum_{j=1}^{j=| \boldsymbol{n}_{\mathsf{v}}(i)|} {r}_{\mathsf{n},\mathsf{v}}(i,j) +\sum_{k=1}^{ k=|\boldsymbol{l}_{\mathsf{v}}(i)|} {r}_{\mathsf{l},\mathsf{v}}(i,k),\\
    &=\sum_{j=1}^{j=| \boldsymbol{n}_{\mathsf{v}}(i)|} {c}_{\mathsf{n},\mathsf{v}}(i,j) +\sum_{k=1}^{ k=|\boldsymbol{l}_{\mathsf{v}}(i)|}  {w}_{\mathsf{l},\mathsf{v}}(i,k),\label{eq4}
\end{aligned}
\end{align}
where Eq.~\ref{eq3} and Eq.~\ref{eq4} denote the resource cost and revenue required for $i$-th VNR embedding, respectively. $\boldsymbol{r}_{\mathsf{n},\mathsf{v}}(i)=\{{r}_{\mathsf{n}, \mathsf{v}}(i,1), {r}_{\mathsf{n}, \mathsf{v}}(i,2), \cdots, {r}_{\mathsf{n}, \mathsf{v}}(i,{|\boldsymbol{n}_{\mathsf{v}}(i)|})\}$, where $|\boldsymbol{n}_{\mathsf{v}}(i)|$ indicates the number of virtual nodes in $i$-th VNR. $\boldsymbol{r}_{\mathsf{l}, \mathsf{v}}(i)=\{{r}_{\mathsf{l}, \mathsf{v}}(i,1), {r}_{\mathsf{l}, \mathsf{v}}(i,2), \cdots, {r}_{\mathsf{l}, \mathsf{v}}(i,{|\boldsymbol{l}_{\mathsf{v}}(i)|})\}$, where $|\boldsymbol{l}_{\mathsf{v}}(i)|$ indicates the number of virtual links in $i$-th VNR. It should be noted that $h$ denotes the path hops of this virtual link in the substrate network. This is mainly due to the fact that the virtual link may span multiple substrate links, and thus its required link resource is equal to the sum of all substrate link resources on the path. In addition, it is important to note that only the successful response of the current VNR produces costs and~revenues.

Moreover, this work adopts evaluation indicators widely used by VNE: long-term average revenue, long-term average revenue-cost ratio, and VNR acceptance success rate, as shown in Eq.~\ref{eq5}, Eq.~\ref{eq6}, and Eq.~\ref{eq7}, respectively.
Among them, the long-term average revenue is expressed as the integral ratio of the revenue to time of all VNRs. The long-run average revenue-cost ratio is expressed as the integral ratio of revenue to cost of all VNRs. The VNR acceptance success rate is expressed as the integral ratio of successfully accepted VNRs to all VNRs. In addition, according to the expression of the equations, it is known that the performance of the algorithm can be reflected by the change of the evaluation indicator, and the larger its value, the better the performance.
% \begin{equation}
%  \begin{aligned}
%         \tilde{\xi}&=\sum^{i=|{\mathsf{V}}|}_{i=1}\left(\lim_{T\rightarrow \infty}\frac{ \int^{t=T}_{t=0} \xi_i(t)\ {d}t}{T}\right)\\
%         &=\sum^{i=|{\mathsf{V}}|}_{i=1}\left(\lim_{
%             \Delta t\rightarrow 0
%         }\frac{ \sum^{\infty}_{n=1} \xi_i(n\Delta t)\ \Delta t}{\sum^{\infty}_{n=1} n\Delta t}\right),  \label{eq5}
%     \end{aligned}
%     \end{equation}
% \begin{equation}
%  \begin{aligned}
%         \eta&=\sum^{i=|{\mathsf{V}}|}_{i=1}\left( \lim_{T\rightarrow \infty}\frac{ \int^{t=T}_{t=0} \xi_i(t)\ {d}t}{ \int^{t=T}_{t=0} \zeta_i(t)\ {d}t}\right)\\
%         &=\sum^{i=|{\mathsf{V}}|}_{i=1} \left(\lim_{\Delta t\rightarrow 0}\frac{ \sum^{\infty}_{n=1} \xi_i(n\Delta t)\  \Delta t}{\sum^{\infty}_{n=1} \zeta_i(n\Delta t)\ \Delta t}\right),\label{eq6}
%     \end{aligned}
% \end{equation}
    \begin{align}
    &\ \ \ \ \ \begin{aligned}
        \tilde{\xi}&=\sum^{i=|{\mathsf{V}}|}_{i=1}\left(\lim_{T\rightarrow \infty}\frac{ \int^{t=T}_{t=0} \xi_i(t)\ {d}t}{T}\right)\\
        &=\sum^{i=|{\mathsf{V}}|}_{i=1}\left(\lim_{
            \Delta t\rightarrow 0
        }\frac{ \sum^{\infty}_{n=1} \xi_i(n\Delta t)\ \Delta t}{\sum^{\infty}_{n=1} n\Delta t}\right),  \label{eq5}
    \end{aligned}\\
     &\ \ \ \ \ \begin{aligned}
        \eta&=\sum^{i=|{\mathsf{V}}|}_{i=1}\left( \lim_{T\rightarrow \infty}\frac{ \int^{t=T}_{t=0} \xi_i(t)\ {d}t}{ \int^{t=T}_{t=0} \zeta_i(t)\ {d}t}\right)\\
        &=\sum^{i=|{\mathsf{V}}|}_{i=1} \left(\lim_{\Delta t\rightarrow 0}\frac{ \sum^{\infty}_{n=1} \xi_i(n\Delta t)\  \Delta t}{\sum^{\infty}_{n=1} \zeta_i(n\Delta t)\ \Delta t}\right),\label{eq6}
    \end{aligned}\\
    &\ \ \ \ \ \begin{aligned}
        \gamma&=\lim_{T\rightarrow \infty} \frac{\int^{t=T}_{t=0}\vartheta(t)\ {d}t}{\int^{t=T}_{t=0}\Theta(t)\ {d}t}\\
        &=\lim_{\Delta t\rightarrow 0} \frac{\sum_{n=1}^{\infty}\vartheta(n\Delta t)\ \Delta t}{\sum_{n=1}^{\infty}\Theta(n\Delta t)\ \Delta t}, \label{eq7}
    \end{aligned}\\
    &\ \ \begin{aligned}
        \vartheta=\sum_{i=1}^{i=|{\mathsf{V}}|}(\prod_{j=1}^{j=|\boldsymbol{n}_{{\mathsf{v}}}(i)|} \alpha_{n}^{n_{{\mathsf{v}}}(i,j)} \times \prod_{k=1}^{k=|\boldsymbol{l}_{{\mathsf{v}}}(i)|} \beta_{l}^{l_{{\mathsf{v}}}(i,k)}),
    \end{aligned}\\
    &\alpha_{n}^{n_{{\mathsf{v}}}(i,j)}=\begin{cases}
    1, & n_{{\mathsf{v}}}(i,j)\  \text{successfully embedded in}\  n,\\
    0, &  \text{others},
    \end{cases}\\
    &\beta_{l}^{l_{{\mathsf{v}}}(i,k)}=\begin{cases}
    1, & l_{{\mathsf{v}}}(i,k)\ \text{successfully embedded in}\  l,\\
    0, &  \text{others},
    \end{cases}
\end{align}
where $\vartheta$ represents the number of successfully accepted VNRs, $\Theta$ represents the number of all VNRs, and $l$ is a collective term for $l_{\mathsf{intra}}$ and $l_{\mathsf{inter}}$.

% In addition, we used two metrics, root mean square error (RMSE) and Pearson correlation coefficient, to measure the accuracy of the VNE strategy, as shown in Eq.~\ref{eq:mse} and Eq.~\ref{eq:pea}. Among them, RMSE is used to measure the degree of deviation between the actual value and the predicted value, with smaller values indicating higher prediction accuracy. Pearson correlation coefficient is used to measure the correlation between two variables, with values closer to $1$ indicating a stronger correlation. 
% \begin{equation}
%     \Phi = \sqrt{\frac{1}{n}\sum_{i=1}^n(y-\hat{y})^2},
%     \label{eq:mse}
% \end{equation}
% \begin{equation}
%     \Psi = \frac{\sum_{i=1}^{n}(y-\bar{y})(\hat{y}-\bar{\hat{y}})}{\sqrt{\sum_{i=1}^{n}(y-\bar{y})^2}\sqrt{\sum_{i=1}^n(\hat{y}-\bar{\hat{y}})^2}},
%     \label{eq:pea}
% \end{equation}
% where $y$ indicates the predicted value, $\hat{y}$ indicates the actual value, $\bar{y}$ indicates the mean of $y$, and $\bar{\hat{y}}$ indicates the mean of~$\hat{y}$.

% r = Σ[(xi-x̄)(yi-ȳ)] / √[Σ(xi-x̄)² * Σ(yi-ȳ)²]
% 其中，xi和yi分别表示两个变量在各个观测点的观测值，x̄和ȳ分别表示两个变量的均值，Σ表示求和。
% 此外，我们采用了均方根误差和皮尔森相关系数两个指标来衡量VNE策略的正确性，如公式10、11所示。其中，均方根误差是用于衡量真实值与预测值之间偏差程度的指标，其值越小说明预测准确度越高。（可能大于1可能小于1，具体与数据集有关）。皮尔森相关系数是一种度量两个变量之间相关性的统计量，其值越接近1说明相关性越强。由于VNE是NP难问题，因此我们采用交叉验证的方法来评估模型的性能。具体地，其将在Section~\ref{sec:decision}中被详细说明。
\section{Algorithm Design of DNFS-VNE}\label{sec3}
Entailment is the core of the fuzzy system, which implies the logical cause-and-effect relationship, expressing reasoning or deductive logic~\cite{10144650}. In traditional fuzzy systems, entailment is realized by fuzzy implication operators, such as \textit{Min} operator, \textit{Max} operator, and so on~\cite{pang2022fuzzy}. It is realized in the form of an affirmative antecedent argument, as follows,
\begin{equation}
    (A=\text{True}) \wedge (A\rightarrow B=\text{True})\Rightarrow B=\text{True},
\end{equation}
It should be noted that assuming that $A'$ is a subset of $A$, the fuzzy implication operator calculates the membership value of $A'$, and the fuzzy rules infer $B'$ according to $A'$, where $B'$ is a subset of $B$. However, for elements that are not in the subset of $A$, it often leads to poor triggering effects of fuzzy rules. Therefore, this traditional fuzzy system is highly dependent on empirical knowledge, which requires training data and fuzzy rules to sufficiently cover possible test scenarios, that is, the reasoning space is required to be large enough. However, this situation is impractical for complex network environments.

Based on the inspiration of DNFS, in order to achieve self-learning, self-adaptation, self-reasoning, \textit{etc.}, of the VNE algorithms, this work aims to use classical DL-model convolutional neural networks (CNNs) to implement the fuzzy implication operator of the traditional fuzzy system and thus realize the implication operation.
It is worth stating that its output will be the membership value of the fuzzy set of the subsequent defuzzification layer.

\vspace{-0.3cm}
\subsection{Model Composition}\label{sec:fiveb}
\begin{figure}
    \centering
    \includegraphics[width=0.75\linewidth]{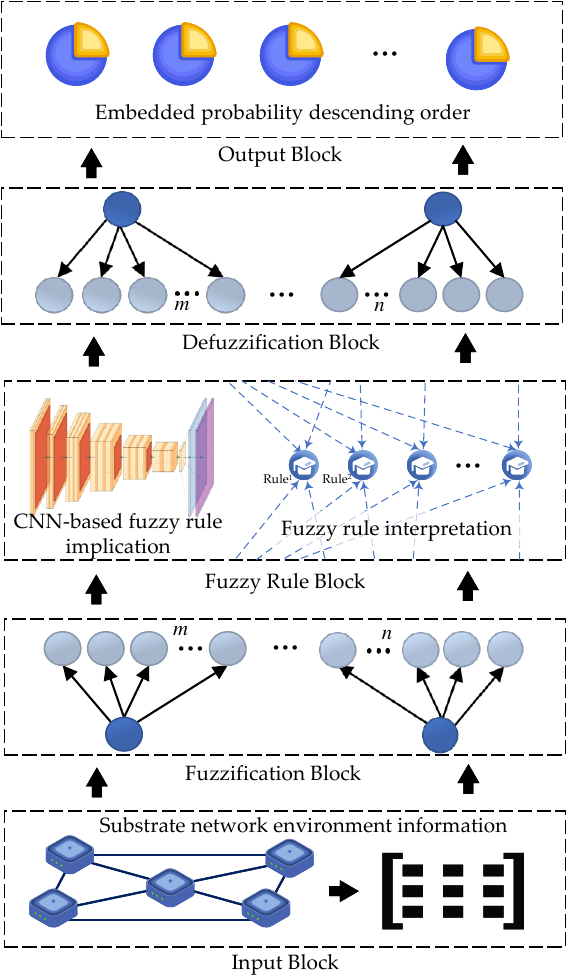}
    \caption{The policy network diagram of the proposed DNFS-VNE algorithm.}
    \vspace{-0.3cm}
    \label{fig:3}
\end{figure} 
Based on the general paradigm of DNFS shown in Fig.~\ref{fig:2}, the policy network diagram of the proposed DNFS-VNE algorithm is shown in Fig.~\ref{fig:3}. Specifically, it also contains five blocks, which are the input block, fuzzification block, fuzzy rule block, defuzzification block, and output block. The specific details of each block are as follows:
\subsubsection{Input block}
This block focuses on extracting the substrate network information and constructing the feature matrix. Same as in previous works~\cite{10132867, cao2017novel, 9894092, yao2020continuous}, the following information is selected as the substrate node feature information:

\textit{Available node resources}: the available node resource of the substrate node $n(i)$ is denoted as,
\begin{equation}
\begin{aligned}
     {r}^{\mathsf{a}}_{\mathsf{n}}(i)&= {r}_{\mathsf{n}}(i)-\sum_{j=1}^{j=|{\mathsf{V}}|}\sum_{k=1}^{k=|\boldsymbol{n}_{{\mathsf{v}}}(j)|}\alpha_{n(i)}^{n_{{\mathsf{v}}}(j,k)}\times {r}_{\mathsf{n},{\mathsf{v}}}(j,k)\\
     &={c}_{\mathsf{n}}(i)-\sum_{j=1}^{j=|{\mathsf{V}}|}\sum_{k=1}^{k=|\boldsymbol{n}_{{\mathsf{v}}}(j)|}\alpha_{n(i)}^{n_{{\mathsf{v}}}(j,k)}\times {c}_{\mathsf{n},{\mathsf{v}}}(j,k),
\end{aligned}
\end{equation}
where $\boldsymbol{r}_{\mathsf{n}}=\{{r}_{\mathsf{n}}(1), {r}_{\mathsf{n}}(2), \cdots, {r}_{\mathsf{n}}({|\boldsymbol{n}|})\}$, $|\boldsymbol{n}|$ indicates the number of substrate nodes.

\textit{Available link resources}: it is expressed as the sum of available bandwidths connected to the substrate nodes. The available link resource of the substrate node $n(i)$ is denoted as,
\begin{equation}
    \begin{aligned}
        {r}^{\mathsf{a}}_{\mathsf{l}}(i)=\sum_{\forall {l({i,j})}} {r}_{\mathsf{l}}(i,j)=\sum_{\forall {l({i,j})}} {w}_{\mathsf{l}}(i,j),
    \end{aligned}
\end{equation}
where $\boldsymbol{r}_{\mathsf{l}}=\{{r}_{\mathsf{l}}(1), {r}_{\mathsf{l}}(2), \cdots, {r}_{{\mathsf{l}}}({|\boldsymbol{l}|})\}$, $|\boldsymbol{l}|$ indicates the number of substrate links, and $l({i,j})$ represents the links between substrate nodes $n(i)$ and $n(j)$.

\textit{Average distance}: it is expressed as the average path length from the substrate node to other nodes. The larger its value, the greater the bandwidth occupied by the links passing through this node. The average distance of the substrate node $n(i)$ to other substrate nodes is denoted as,
\begin{equation}
    {d}(i)=\frac{ \sum_{\forall {l({i,j})}} \|{l}(i)-{l}(j) \|_2 }{1+h({i,j})},
\end{equation}
where ${l}(i)$ denotes the location of the substrate node $n(i)$, ${l}(j)$ denotes the location of the substrate node $n(j)$, $h(i,j)$ denotes the hops of the links between substrate nodes $n(i)$ and $n(j)$, and $\|$ denotes the Euclidean distance.

To eliminate the scale difference between various features, the input data is normalized using max-min normalization. This process maps the different data ranges into a uniform scale range, as follows,
\begin{equation}
\hat{x}=\frac{x-\text{min}(\boldsymbol{x})}{\text{max}(\boldsymbol{x})-\text{min}(\boldsymbol{x})},
\end{equation}
where $\boldsymbol{x}=[x_1,x_2,\cdots, x_{|\boldsymbol{n}|}]^{\text{T}}$, $\text{min}()$ represents the minimum value of the vector, and $\text{max}()$ represents the maximum value of the vector.

Thus, the input information is matrixed as follows,
\begin{equation}
\begin{aligned}
             \mathbf{X}=
             \begin{bmatrix}\hat{\boldsymbol{x}}_1\\ \hat{\boldsymbol{x}}_2\\ \vdots\\ \hat{\boldsymbol{x}}_{|\boldsymbol{n}|}  \end{bmatrix}
    =\begin{bmatrix}
   \hat{r}_{\mathsf{n}}^{\mathsf{a}}(1)  & \hat{r}_{\mathsf{l}}^{\mathsf{a}}(1)  & \hat{{d}}(1) \\
   \hat{r}_{\mathsf{n}}^{\mathsf{a}}(2)  & \hat{r}_{\mathsf{l}}^{\mathsf{a}}(2)  & \hat{{d}}(2) \\
    \vdots & \vdots & \vdots\\
    \hat{r}_{\mathsf{n}}^{\mathsf{a}}(|\boldsymbol{n}|)  & \hat{r}_{\mathsf{l}}^{\mathsf{a}}(|\boldsymbol{n}|)  & \hat{{d}}(|\boldsymbol{n}|) \\
    \end{bmatrix}.
\end{aligned}
\label{eq14}
\end{equation}

\subsubsection{Fuzzification block}
This block involves fuzzification of the input data through a Gaussian membership function, which converts it into a fuzzy representation. The calculation is based on the distance between the input data $x$ and the center point $c$, where smaller distances result in larger membership values. As the distance increases, the membership value decreases until it reaches $0$. Here are the specifics,
\begin{equation}
    \mathcal{F}(x)=\exp\left\{-\frac{(x-c)^2}{2\sigma^2}\right\},
    \label{eq15}
\end{equation}
where $\mathcal{F}(x)$ denotes the membership value corresponding to the input value $x$, $c$ denotes the centroid of the membership function, and $\sigma$ denotes the width of the membership function. In this work, we utilize the clustering algorithm~\cite{likas2003global} to determine the centroids of clusters as $c$ and utilize the standard deviation as $\sigma$.

It should be noted that, as described in Section~\ref{sec2c}, the fuzzy representation represents a method for processing uncertain information, including fuzzy sets and membership values. Wherein, the fuzzy set is utilized to express and process imprecise information in a fuzzy system. The uncertain state of the system is abstracted into a fuzzy set, which enables the data to be expressed in a linguistic manner.
Moreover, the degree of affiliation of the data belonging to the corresponding fuzzy set is expressed through the membership value, which is a real number between $0$ and $1$. For example, setting three membership functions for the variable $x$ will generate the fuzzy representation containing three fuzzy linguistic labels and corresponding membership values (\textit{e.g.}, ``Large'' = $0.82$, ``Medium'' = $0.19$, ``Small'' = $0$).
\subsubsection{Fuzzy rule block}
This block consists of two branches, the CNN-based fuzzy implication branch and the fuzzy rule interpretation branch. In this work, the fuzzy implication branch is a CNN-based implication operator that outputs the inferred membership value, which serves as the input to the defuzzification block, and undergoes aggregation and defuzzification to obtain the output substrate network node embedding probability. In addition, the fuzzy rule interpretation branch establishes associations between the fuzzy sets with the highest membership values from the fuzzification block and the defuzzification block. These associations form fuzzy rules that can be interpreted based on their linguistic labels, thereby breaking away from the traditional black-box paradigm of DL models. In other words, these rules act as a mapping between input to output spaces.

\textbf{CNN-based fuzzy rule implication branch}:
This branch is primarily composed of two convolutional layers and two fully connected layers to compute the available computational vectors of the substrate network and infer the membership values of the consequent fuzzy rules as the input to the defuzzification layer. The convolutional layer is computed as follows:
\begin{equation}
\vspace{-0.1cm}
    \mathbf{C}=Conv(\mathbf{{F}}, \mathbf{{W}})+\boldsymbol{b},
    \vspace{-0.1cm}
\end{equation}
where $\mathbf{F}$ represents the output of the fuzzification block, $\mathbf{W}$ represents the weight matrix of the convolution layer, and $\boldsymbol{b}$ represents the bias vector. Moreover, the weights are initially initialized using the truncated normal distribution method to avoid the vanishing or exploding of the gradient caused by extreme values. This method restricts $99\%$ of the weights to a range between $-3\sigma$ and $3\sigma$. In addition, the last fully connected layer is responsible for producing the membership values of the consequent, with the number of neurons equal to $q$. The optimizer uses a gradient descent optimizer to minimize the loss function through forward propagation and gradient back-propagation (BP) of the model. During this process, the model will identify fuzzy rule patterns and cache them in weights. It is worth explaining that the input and output of the CNN-based fuzzy implication branch are associated with the subsequent fuzzy rule interpretation branch, which contributes to the interpretability of rule implication.

% In addition, the loss function of this work is cross-entropy loss, as follows,
% \begin{equation}
%     \mathcal{L}=log().
% \end{equation} 损失函数不在这，在最后
% 嵌入式CNN模型的前向和后向学习和自适应是为了识别神经模糊系统的蕴涵模式并将其缓存为核权重。

% 在接收到测试数据的模糊表示后，深度学习蕴涵检索权重并按照与识别的模糊蕴涵模式相同的蕴涵方式计算输出隶属值。这提供了一种预测可能与训练数据显着不同的测试数据的解决方案。同时，深度学习模糊蕴涵的蕴涵遵循从训练数据中提取的知识。此外，它不受传统神经模糊系统中规则剪枝过程的限制，传统神经模糊系统可能会因激发强度低而删除关键信息。深度学习结构保留了训练期间识别的所有信息，并且只允许数据定义蕴涵运算。

% 将深度学习模型嵌入神经模糊架构的优点在于，它允许数据驱动的模糊蕴涵提供与现实世界数据蕴含的密切对应，使整个框架对于每个单独的预测都是可解释的。最重要的是，Mamdani 模糊神经模型常见的模糊语言标签也与深度学习模型的输入和输出相关联。这种推理机制使用相同的语义集赋予深度学习含义透明性。

%这里不严谨，输入数据指的是模糊规则解释分支，而不是最初的input data
\textbf{Fuzzy rule interpretation branch}: Fuzzy rules are based on the more intuitive and interpretable Mamdani-type\footnote{It is an antecedent-consequent type rule and uses linguistic labels in both antecedent and consequent parts.} linguistic rule~\cite{iancu2012mamdani}. In this work, the $j$-th input data is $\boldsymbol{x}_j=\left[x_1, x_2, x_3\right]$ ($1\leq j\leq |\boldsymbol{n}|$) as in Eq.~\ref{eq14}. The fuzzy set size\footnote{In general, the more Gaussian membership functions there are, the more precise the representation of the fuzzy set becomes. However, this may also lead to increased computation complexity and decreased operational efficiency. Hence, when determining the number of Gaussian membership functions, trade-offs and choices need to be made based on actual needs and computing resources.} (number of fuzzy labels, depending on the number of membership functions) is defined as $5$, \textit{i.e.}, linguistic labels ``Very High (VH)'', ``High (H)'', ``Medium (M)'', ``Low (L)'', and ``Very Low (VL)''. Therefore, the encoding of fuzzy rules derived from the fuzzification layer (antecedent) and defuzzification layer (consequent) is as follows,
\begin{equation}
\begin{aligned}
    \mathbb{R}^i:\  &\textbf{If}\  x_1=F^i_1,\ \text{and}\ x_2=F^i_2,\ \text{and}\ x_3=F^i_3, \\
    &\textbf{Then}\  y_1=D^i_1,\ \text{and}\ y_2=D^i_2,\cdots,\ \text{and}\ y_q=D^i_q,
\end{aligned}
\label{eq16}
\end{equation}
where $\mathbb{R}^i$ represents the $i$-the fuzzy rule, $F_k^i$ represents the linguistic label corresponding to the $k$-th input dimension of $\mathbf{X}$ from the fuzzification black, and $D^i_l$ represents the linguistic label corresponding to the $l$-th output dimension from the defuzzification black.
% i表示第i条规则

\subsubsection{Defuzzification block}
This block uses the center of gravity method for defuzzification to calculate the weighted average of the membership values of the fuzzy output and then uses the center of gravity as the output value. Unlike the maximum method, which may ignore relatively small membership values that have a certain impact, this method considers the possible impact of all membership values. Specifically, for the substrate node $n(i)$, it is calculated as follows,
\begin{equation}
    {o}(i)=\frac{{f}(i)\times {\mathcal{F}}'(i)}{\sum_{j=1}{\mathcal{F}}'(j)},
    \label{eq18}
\end{equation}
where $o(i)$ represents the center of gravity of the substrate node $n(i)$, and ${\mathcal{F}}'(i)$ represents the membership value of the $i$-th dimension output ${f}(i)$ of the last fully connected layer at the CNN-based fuzzy rule implication branch.

\subsubsection{Output block}
This block utilizes the softmax function to output the embedded probabilities of the substrate network nodes. Specifically, for the substrate node $n(i)$, it is calculated as follows,
\begin{equation}
    {p}(i)=\frac{{o}(i)}{\sum_{j=1}^{|\boldsymbol{n}|} {o}(j)},
    \label{eq19}
\end{equation}
where $p(i)$ represents the embedded probability of $n(i)$.

Finally, after sorting, a set of candidate physical nodes with embedding probabilities from large to small is obtained. To further clarify the fuzzy rule implication process, we connect the consequent of the fuzzy rule interpretation branch to the output block. Thus, it can be further represented as,
\begin{equation}
\begin{aligned}
    \mathbb{R}^i:\  &\textbf{If}\  x_1=F^i_1,\ \text{and}\ x_2=F^i_2,\ \text{and}\ x_3=F^i_3, \\
    &\textbf{Then}\  F'^i_j:w^i_j=p_j,
\end{aligned}
\label{eqa16}
\end{equation}
where $F'^i_j$ represent the linguistic label of $j$ substrate node, $w^i_j$ represent the rule weight of $\mathbb{R}^i$, $1\leq j\leq {|\boldsymbol{n}|}$, and $p_{j}$ represents a specific probability value.

The process of establishing the rule base is as follows: the establishment of rules is data-driven. When the first data is input, based on Eq.~\ref{eqa16}, the first rule is also established. If existing rules cannot explain current actions, new rules will be created. These rules are combined into a rule base to explain the principles and meaning of CNN-based fuzzy implication.
\subsection{DRL Configuration}
The basic elements of DRL for this work are configured as follows,

\subsubsection{Agent} The main entity of DRL, which performs actions in the environment and learns how to make optimal decisions through interaction with the environment. Its configuration is shown in Fig.~\ref{fig:3}, and detailed information is provided in Section~\ref{sec:fiveb}.

\subsubsection{State} It refers to the information collection of the environment at a certain moment, as defined in Eq.~\ref{eq14}, and is also the input information extracted by the input block.

\subsubsection{Action} It refers to the scheme adopted by the agent at a certain moment. It is defined as the node embedding scheme and the link embedding scheme, as shown in Eq.~\ref{eq:action}. It should be noted that the node embedding probability is obtained from the output block, and the link embedding scheme is obtained through breadth-first search (BFS). At time $t$, for $\mathcal{V}_i$, the action $a_{i}(t)$ adopted is defined as Eq.~\ref{eq:action}, where $a_{\mathsf{n}}(t, j)=1$ indicates that the substrate node $n(j)$ is used to host the virtual node, and $a_{\mathsf{l}}(t, j)=1$ indicates that the substrate link $l(k)$ is used to host the virtual link. Furthermore, the mapping relationship between the virtual node and the substrate node is $1:1$, so $\sum_{j=1}^{|\boldsymbol{n}|}\ a_{\mathsf{n}}(t, j)=| \boldsymbol{n}_{\mathsf{v}}(i)|$. However, the mapping relationship between virtual links and physical links is $1:m$, which means that the virtual link may span multiple substrate links (also mentioned in Eq.~\ref{eq3}).
\begin{figure*}
\begin{equation}
\begin{aligned}
    a_{i}(t)=\Bigg\{
    \bigg\{
    \Big(a_{\mathsf{n}}(t, 1), a_{\mathsf{n}}(t, 2),\cdots,a_{\mathsf{n}}(t, j),\cdots, a_{\mathsf{n}}(t,|\boldsymbol{n}|)\Big)\ &\Big|\ a_{\mathsf{n}}(t, j)=0,1\ \text{and}\sum_{j=1}^{|\boldsymbol{n}|}\ a_{\mathsf{n}}(t, j)=| \boldsymbol{n}_{\mathsf{v}}(i)|
    \bigg\},\\
    \bigg\{
    \Big(a_{\mathsf{l}}(t, 1), a_{\mathsf{l}}(t, 2),\cdots,a_{\mathsf{l}}(t, k),\cdots, a_{\mathsf{l}}(t,|\boldsymbol{l}|)\Big)\ &\Big|\ a_{\mathsf{l}}(t, k)=0,1
    \bigg\}
    \Bigg\},
    \end{aligned}
    \label{eq:action}
\end{equation}
 \hrulefill
\end{figure*}

\subsubsection{Reward}It refers to the feedback signal value obtained by an action taken by the agent at a certain moment, which is used to guide the agent to continue learning in the direction of a greater positive feedback value. Moreover, it is defined as the long-term average revenue-cost ratio, as follows,
\begin{equation}
    r(t)=\eta.
    \label{eq:reward}
\end{equation}

\begin{algorithm}[]
\small
 \caption{{The Learning Process of DNFS-VNE}}
 \label{alg:algorithm1}
 
 \KwIn{{Substrate network and VNRs.}}
 \KwOut{Indicators Eq.~\ref{eq5}, Eq.~\ref{eq6}, Eq.~\ref{eq7}; Fuzzy rule base $\mathbf{R}$.}
 \BlankLine
 {Truncated normal distribution randomly initializes network weights;  $\mathbf{R}={\varnothing}$.}
 
 \While{{$iteration\leq max\_iteration $}}{
  \ForEach{{$n(i)\in\boldsymbol{n}$}}{
  {Build the state by Eq.~\ref{eq14};}
  
   Fuzzify input information by Eq.~\ref{eq15};

   {CNN-based fuzzy rule implication branch forward propagation;}

Defuzzification by Eq.~\ref{eq18};

Calculate node embedding probability by Eq.~\ref{eq19};

   \eIf{$n_{\mathsf{v}}(i,j)$ embedded is succeed}{
   Search substrate links by BFS;
   
   \eIf{$l_{\mathsf{v}}(i,k)$ embedded is succeed}{
   Update $\mathbf{R}$ by Eq.~\ref{eqa16};

Calculate reward value by Eq.~\ref{eq:reward};
   
   Calculate loss value by Eq.~\ref{eq20};
   
   Calculate gradient by Eq.~\ref{eq:21} and update parameters via BP;
   }{Cancel gradient BP;}
   }{Cancel gradient BP;}
   }
  {$iteration=iteration+1$;}
 }
\end{algorithm}
\vspace{-0.2cm}
\subsection{Model Learning}
As mentioned earlier, DNFS-VNE utilizes the BP algorithm to update the identified implication patterns into the weights. Specifically, the loss function employed in this work is the cross-entropy loss, as follows,
\begin{equation}
    \mathcal{L}=- \sum_i^{|\boldsymbol{n}|}{o}(i)\log({p}(i)).
    \label{eq20}
\end{equation}

Furthermore, the gradient is updated in the following directions,
\begin{equation}
    \mathcal{L}:=\mathcal{L}-\mu\times r(t)\times \mathcal{L}',
    \label{eq:21}
\end{equation}
where $\mu$ represents the learning rate and $\mathcal{L}'$ represents the gradient derivative of $\mathcal{L}$. It should be noted that the reward $r(t)$ will be used to jointly guide the model to optimize towards high reward feedback signals. It is widely recognized that the learning rate is a crucial parameter that requires careful tuning. Setting the learning rate too high or too low can hinder the model's convergence. In this work, the learning rate is confirmed to be $0.01$ after being verified in Section~\ref{sec4b}. During the learning process, the model will be optimized along the direction of gradient descent, following Eq.~\ref{eq:21}, until the loss function reaches a converged state. At this point, the three evaluation indicators as shown in Eq.~\ref{eq5}, Eq.~\ref{eq6}, and Eq.~\ref{eq7} will also converge.

The algorithm flow of the proposed DNFS-VNE is shown in Algorithm~\ref{alg:algorithm1}. It should be noted that the current VNR $R_i$ embedding is successful only when all virtual nodes $n_{\mathsf{v}}(i,j)(1\leq j\leq |\boldsymbol{n}_{\mathsf{v}}(i)|)$ and virtual links $l_{\mathsf{v}}(i,k)(1\leq k\leq |\boldsymbol{l}_{\mathsf{v}}(i)|)$ are successfully embedded. At this point, the learned rule patterns are updated to the model weights for the current successful embedding. Otherwise, no learning is done for the failed embedding.

In addition, the rule base is built through inference in traditional fuzzy systems. This process involves pruning fuzzy rules with weak firing strength, less coverage, and less weight in order to ensure compactness. However, this approach may result in the removal of key information, as the pruned rules often imply that the event occurrences are relatively rare. 
On the other hand, DNFS-VNE can disregard this aspect, which is due to that the rule base is built by extracting knowledge from the data and implementing it based on the computation of DNNs. Then, the identified fuzzy rule patterns are updated into the weights. As a result, DNFS-VNE is not bound by rule pruning and does not require the training data to sufficiently cover the test data, as is the case in traditional fuzzy systems.
% In addition, the loss function of this work is cross-entropy loss, as follows,
% \begin{equation}
%     \mathcal{L}=log().
% \end{equation} 损失函数不在这，在最后
% 嵌入式CNN模型的前向和后向学习和自适应是为了识别神经模糊系统的蕴涵模式并将其缓存为核权重。

% 在接收到测试数据的模糊表示后，深度学习蕴涵检索权重并按照与识别的模糊蕴涵模式相同的蕴涵方式计算输出隶属值。这提供了一种预测可能与训练数据显着不同的测试数据的解决方案。同时，深度学习模糊蕴涵的蕴涵遵循从训练数据中提取的知识。此外，它不受传统神经模糊系统中规则剪枝过程的限制，传统神经模糊系统可能会因激发强度低而删除关键信息。深度学习结构保留了训练期间识别的所有信息，并且只允许数据定义蕴涵运算。

% 将深度学习模型嵌入神经模糊架构的优点在于，它允许数据驱动的模糊蕴涵提供与现实世界数据蕴含的密切对应，使整个框架对于每个单独的预测都是可解释的。最重要的是，Mamdani 模糊神经模型常见的模糊语言标签也与深度学习模型的输入和输出相关联。这种推理机制使用相同的语义集赋予深度学习含义透明性。
\section{Simulation Experimental Verification and Analysis}\label{sec4}
\subsection{Environment Configuration}
% 在这项工作中，环境配置是通过GT-ITM工具生成并保存于``.txt''文件中。具体地，总共生成100个基板节点和600条基板链路。基板结点随机分布于4个物理域。VNR生成2000个，并记录于2000个``reqnum.txt''中，并且虚拟链路是按照一半的概率随机生成的。此外，所有VNRs按照泊松分布的时间序列到达DNFS-VNE以生成一个连续的过程。
\begin{table}[!htbp]
\caption{The environment configurations of this work.}
    \label{tab:conf}
\centering
\renewcommand\arraystretch{1.3}
\begin{threeparttable}
\begin{tabular}{|l|l||l|l|}
\hline
\textbf{Substrate Network} & \textbf{Setting} & \textbf{Virtual Network }& \textbf{Setting} \\ \hline
Substrate domain & $4$ & VNRs & $2000$\\ 
$\boldsymbol{n}$ & $100$ &  Training or Testing & $1000$\\ 
$\boldsymbol{l}_{\mathsf{intra}}$ and $\boldsymbol{l}_{\mathsf{inter}}$ & $600$ &   $\boldsymbol{n}_{\mathsf{v}}$& $2$-$10$\\ 
$\boldsymbol{r}_{\mathsf{n}}$ & U{[}$50,100${]}\tnote{\textcolor{red}{*}}& $\boldsymbol{r}_{\mathsf{n},\mathsf{v}}$ & U{[}$1,50${]} \\ 
$\boldsymbol{r}_{\mathsf{l}}$ & U{[}$50,100${]} &  $\boldsymbol{r}_{\mathsf{l},\mathsf{v}}$& U{[}$1,50${]} \\ 
\hline 
\end{tabular}
\begin{tablenotes}
        \footnotesize
        \item[*] This indicates that the range of values for all elements in $\boldsymbol{r}_{\mathsf{n}}$. The similar as $\boldsymbol{r}_{\mathsf{l}}$, $\boldsymbol{r}_{\mathsf{n},\mathsf{v}}$, and $\boldsymbol{r}_{\mathsf{l}, \mathsf{v}}$.
      \end{tablenotes}
    \end{threeparttable}

\end{table}
In this work, as shown in Table~\ref{tab:conf}, the environment settings are generated by the GT-ITM tool and saved in ``.txt'' files. Specifically, a total of $100$ substrate nodes and $600$ substrate links are generated. These substrate nodes are randomly distributed in $4$ substrate domains. Additionally, $2,000$ VNRs are generated and recorded in $2,000$ ``.txt'' files, with half as a training set and half as a test set. In each VNR, there are $2$-$10$ virtual nodes, and virtual links are randomly generated with half probability, so there are $\frac{|\boldsymbol{n}_{\mathsf{v}}|\times(|\boldsymbol{n}_{\mathsf{v}}|-1)}{4}$ virtual links. Furthermore, all VNRs arrive at the DNFS-VNE according to a Poisson-distributed time series to generate a continuous~process.

\begin{figure}
    \centering
    \includegraphics[width=\linewidth]{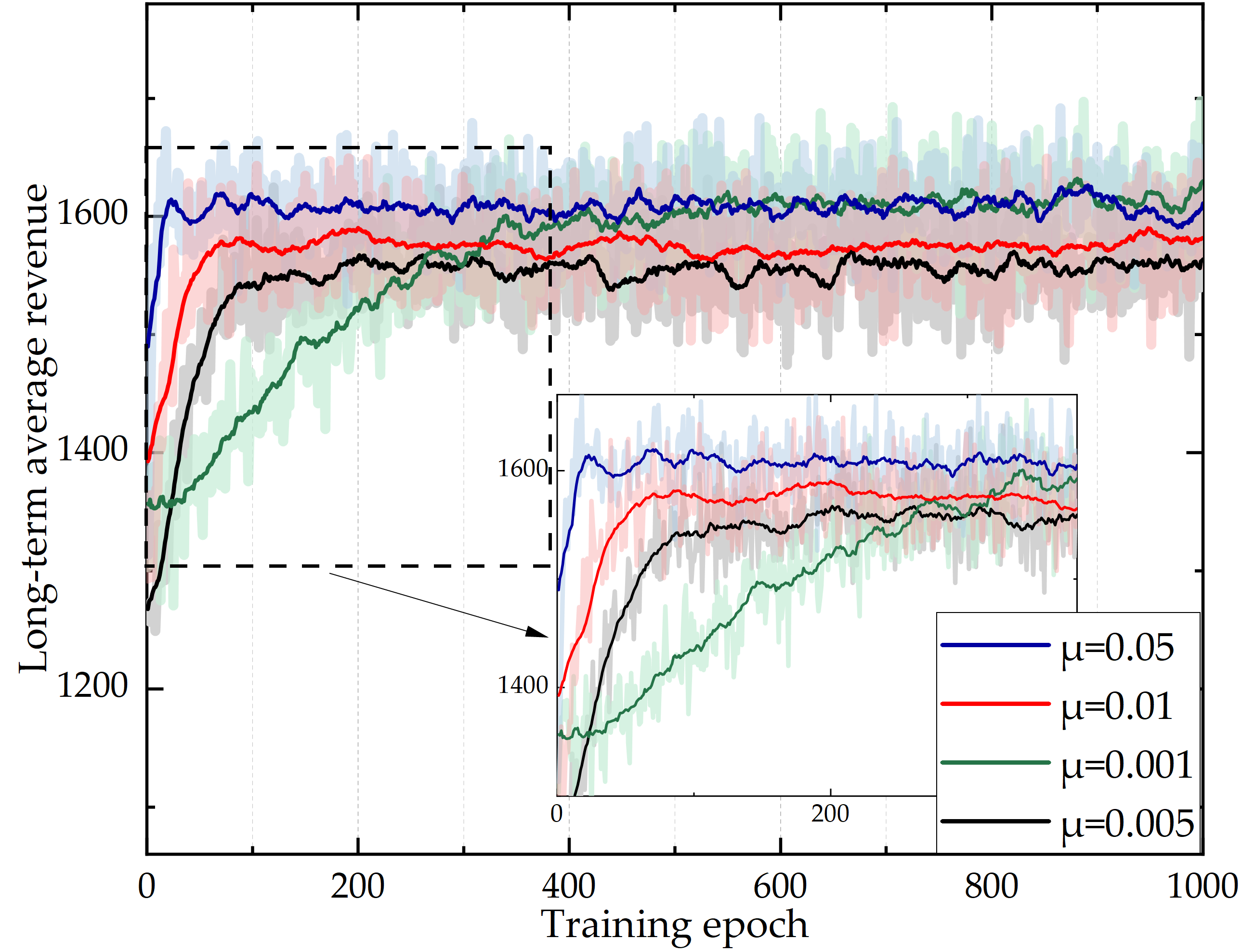}
    \caption{Eq.~\ref{eq5} in the training process.}
    \label{fig:4}
\end{figure}
\subsection{Learning Rate Selection and Training Performance}\label{sec4b} %0.001 0.01 0.05 0.005
\begin{figure}
    \centering
    \includegraphics[width=\linewidth]{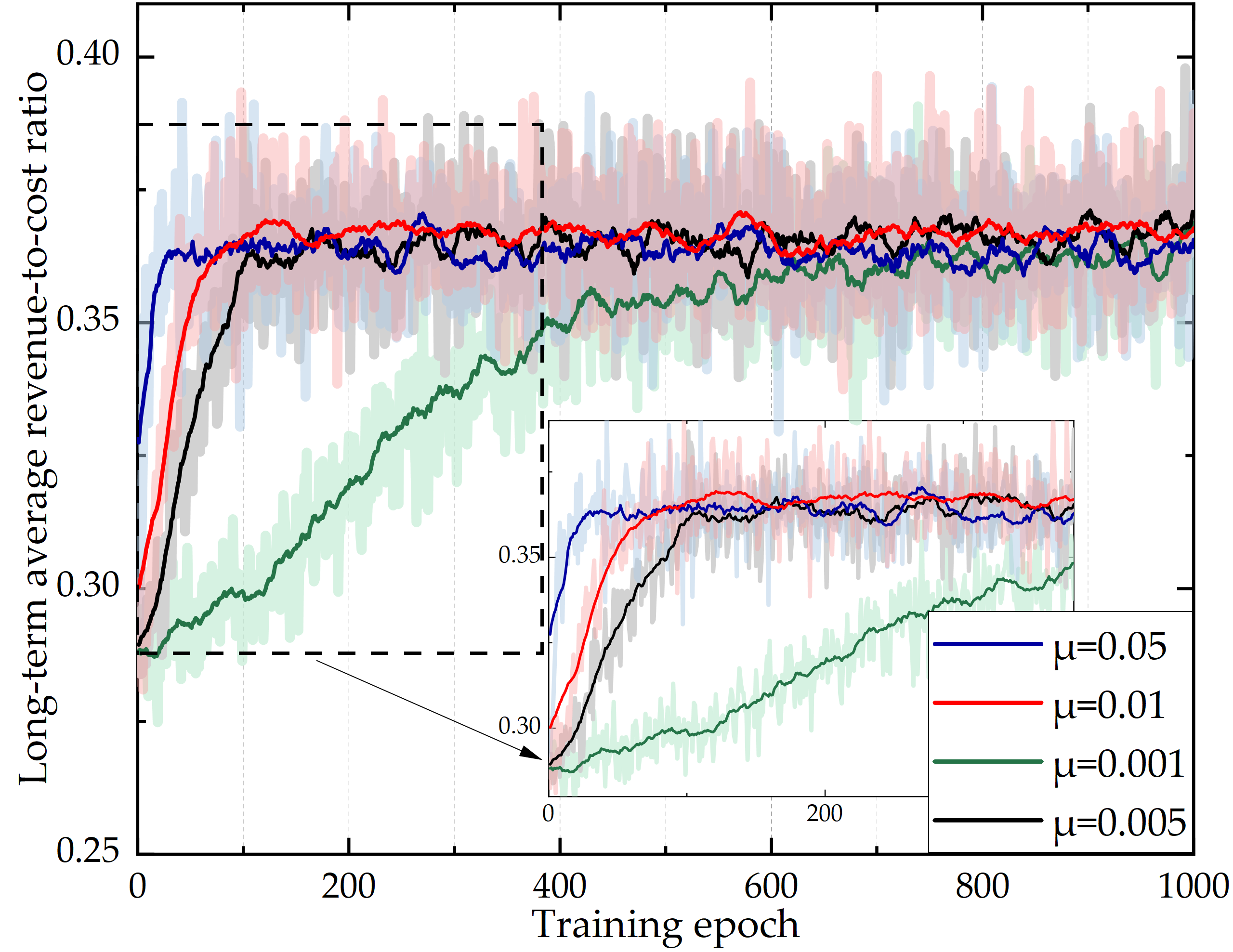}
    \caption{Eq.~\ref{eq6} in the training process.}
    \label{fig:5}
\end{figure}
\begin{figure}
    \centering
    \includegraphics[width=\linewidth]{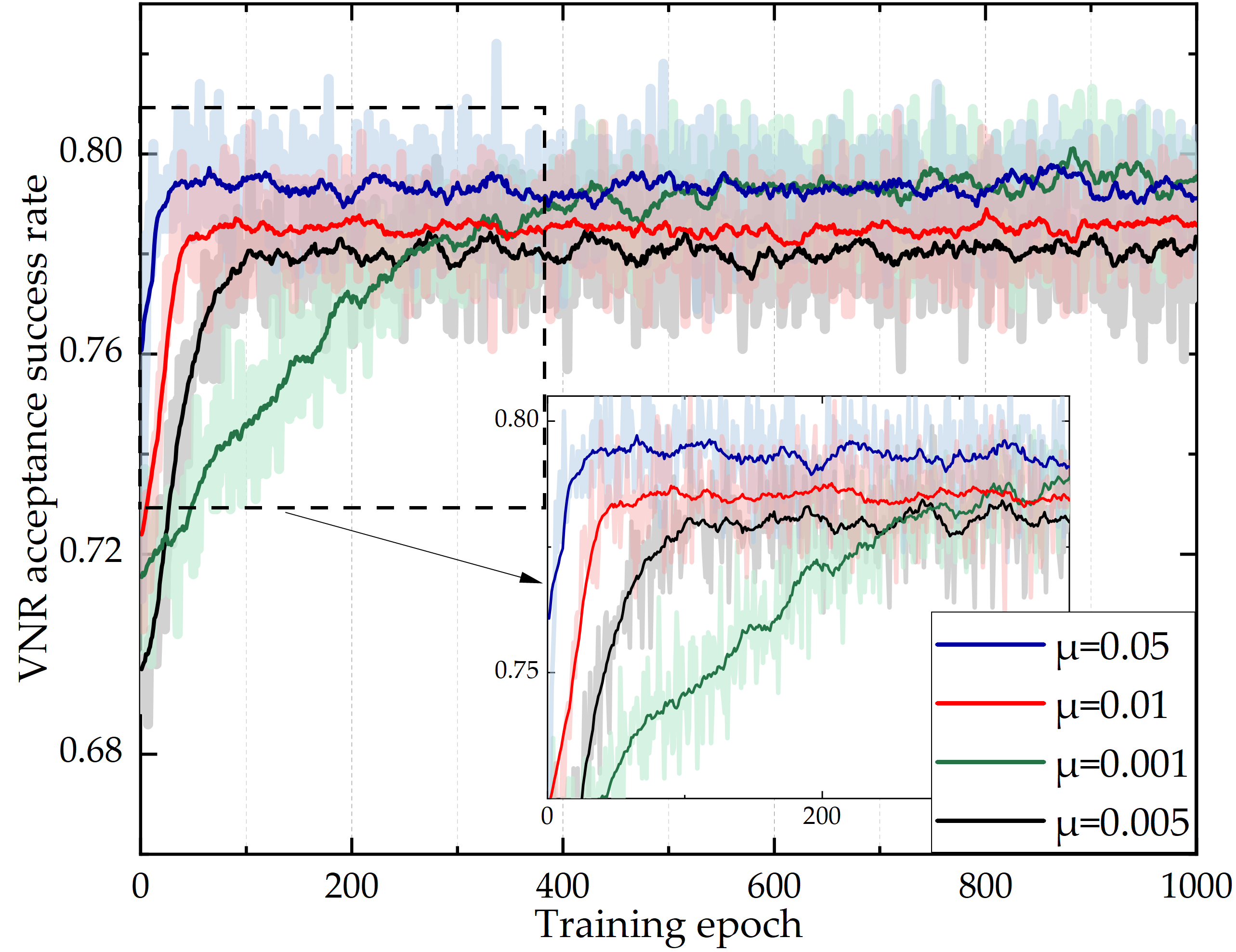}
    \caption{Eq.~\ref{eq7} in the training process.}
    \label{fig:6}
\end{figure}

To determine the learning rate $\mu$ selection and stability of DNFS-VNE, it is important to observe the changes in indicator values at different learning rates. Therefore, we set $\mu$ to $0.01$, $0.05$, $0.005$, and $0.001$, and demonstrate the corresponding changes in indicator values during the training phase in Fig.~\ref{fig:4}, Fig.~\ref{fig:5}, and Fig.~\ref{fig:6}, respectively. It is observed that with $\mu=0.001$, the stability speed is very slow. With $\mu=0.005$, although stability can be achieved, the speed is unacceptable. Moreover, with $\mu=0.05$, stability is faster, but the indicator fluctuates greatly due to the large stride. In contrast, when $\mu=0.01$, due to the appropriate stride, the stability speed is faster and more stable, and it is stable in approximately $30$-$40$ iterations.
Therefore, considering overall performance, stability, and speed, we set $\mu=0.01$.
In addition, with $\mu=0.01$, it can be observed that each evaluation indicator can reach the stability state quickly and more stably during the training~process.

\subsection{Fuzzy Linguistic Rules Definition}
\begin{figure}[!htbp]
    \centering
    \includegraphics[width=0.9\linewidth]{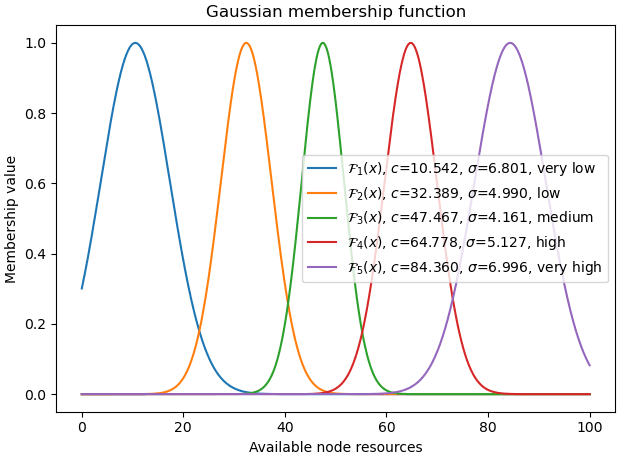}
    \caption{Fuzzification of the available node resource dimension.}
    \label{fig:gauss}
\end{figure}
During the learning process, we apply fuzzification to each input dimension of $\mathbf{X}$ and use the clustering algorithm to determine the centroid and standard deviation of the Gaussian affiliation function. And, we randomly selected a substrate node from one iteration and visualized its fuzzification processing on the dimension of ``Available node resources'', as shown in Fig.~\ref{fig:gauss}. As mentioned earlier, we generate five clusters that correspond to the linguistic labels ``Very High'', ``High'', ``Medium'', ``Low'', and ``Very Low''.  By analyzing the membership values of the current dimension values, we can determine the contribution values of the current input dimension values to different linguistic labels, which are then assigned to the corresponding fuzzy partitions. For instance, if ${r}^{\mathsf{a}}_{\mathsf{n}}(i)=30$, $\mathcal{F}_2({r}^{\mathsf{a}}_{\mathsf{n}}(i))>\mathcal{F}_1({r}^{\mathsf{a}}_{\mathsf{n}}(i))>\mathcal{F}_3({r}^{\mathsf{a}}_{\mathsf{n}}(i))>\mathcal{F}_4({r}^{\mathsf{a}}_{\mathsf{n}}(i))>\mathcal{F}_5({r}^{\mathsf{a}}_{\mathsf{n}}(i))$, we define that during this iteration process, the linguistic interpretation for this substrate node is ``${r}^{\mathsf{a}}_{\mathsf{n}}(i)\ \text{is low}$''. Therefore, based on Mamdani-type linguistic rules and following the same linguistic rules definition at the antecedent and consequent layers, reasonable interpretability can be provided for the VNE algorithm.

\subsection{Generated Fuzzy Rule Interpretation}
We show some generated rules by the fuzzy rule interpretation branch to explore the principles of the embedding process of the VNE, as shown in Table~\ref{tab:result},
\begin{table}[!htbp]
    \centering
        \caption{Some examples of generated fuzzy rules.}
    \label{tab:result}
    \renewcommand\arraystretch{1.5}
    \begin{tabular}{|p{5.5cm}|}
    \hline
    $\begin{aligned}
        \mathbb{R}^1:\  &\textbf{If}\  x_1\ \text{is very high (VH)},\\
        &\quad \text{and}\ x_2\ \text{is very high (VH)},\\
         &\quad \text{and}\ x_3\ \text{is low (L),} \\
    &\textbf{Then}\  ned\  \text{is very high (VH)}: 0.9170,
    \end{aligned}$
    \\ \hline
    $\begin{aligned}
        \mathbb{R}^{2}:\  &\textbf{If}\  x_1\ \text{is low (L)},\\
        &\quad \text{and}\ x_2\ \text{is very high (VH)},\\
         &\quad \text{and}\ x_3\ \text{is low (L),} \\
    &\textbf{Then}\  ned\  \text{is high (H)}: 0.8384,
    \end{aligned}$
    \\ \hline
    $\begin{aligned}
        \mathbb{R}^{3}:\  &\textbf{If}\  x_1\ \text{is very high (VH)},\\
        &\quad \text{and}\ x_2\ \text{is low (L)},\\
         &\quad \text{and}\ x_3\ \text{is low (L),} \\
    &\textbf{Then}\  ned\  \text{is medium (M)}: 0.5916,
    \end{aligned}$
    \\ \hline
   $ \begin{aligned}
        \mathbb{R}^{4}:\  &\textbf{If}\  x_1\ \text{is very high (VH)},\\
        &\quad \text{and}\ x_2\ \text{is low (L)},\\
         &\quad \text{and}\ x_3\ \text{is very high (VH),} \\
    &\textbf{Then}\  ned\  \text{is low (L)}: 0.3811,
    \end{aligned}$
    \\ \hline
    \end{tabular}
\end{table}
where $ ned$ represents the substrate node embedded probability. As mentioned before, $x_1$, $x_2$, $x_3$ represent available node resources, available link resources, and average distance, respectively. Therefore, by observing the comparison between different fuzzy rules, we can explain the embedding principle of the VNE process based on DNNs.
For instance, by comparing $\mathbb{R}^1$ and $\mathbb{R}^{2}$, we can conclude that the more remaining resources in the substrate node, the easier it is to be embedded. Similarly, by comparing $\mathbb{R}^1$ and $\mathbb{R}^{3}$, we can conclude that the more available bandwidth resources in the substrate link, the easier it is to be embedded. Furthermore, by comparing $\mathbb{R}^{3}$ with $\mathbb{R}^{4}$, we can conclude that the lower the number of substrate paths, the higher the embedding revenue. In fact, these reasoning results also align with the original intention of attribute design in many VNE works~\cite{10132867, 10132506, cao2017novel, 9894092, yao2020continuous}. Therefore, this work provides a promising and interpretable solution to the black-box principle explanation of VNE working.

% 此外，需要说明的是，这项工作仅仅是利用DNFS对VNE算法可解释性的第一次尝试。通过这项工作可以很容易的推广到更多的资源维度，例如基板链路延迟、基板结点的度等等，以探索更多属性对VNE过程的影响。也可以探索更细致的模糊推理设计，以获取更细致的模糊规则，进而开发DNFS-VNE更多的版本。

% 在学习过程中，我们X的每个输入维度进行模糊化，并采用聚类算法确定高斯隶属函数的的质心和标准差。我们随机选取了一次迭代的某个基板节点，并针对其“节点可用”这一维度资源的模糊处理进行了可视化，如图7所示。如前所述，我们主要生成5个聚类，分别表示语言标签为A、B、C。根据当前维度值的隶属度值，可以确定当前输入维度值对于不同语义标签的贡献值,进而分配到相应的模糊分区。例如，当a=30时，F2>F1>F3>F4>F5,因此我们定义在此次迭代过程中，对于这个结点来说，其语义解释为:a=F。因此，基于Mamdani型语言规则，在前件层与后件层按照相同的语言规则定义，可以为VNE算法提供一种合理的可解释性。
\subsection{Comparison Experiment Analysis}
\begin{table}[!htbp]
    \centering
        \caption{\textcolor{black}{The descriptions and parameter settings of all baselines.}}
    \label{tab:para}
    \renewcommand\arraystretch{1.3}
    \begin{tabular}{|p{1.5cm}|p{3.5cm}|p{2.2cm}|}
         \hline 
         Baseline & Descriptions & Parameter Settings \\ \hline
         {NodeRank \cite{cheng2011virtual}} & A heuristic algorithm based on the priority of node resources. & $\epsilon=0.0001$, $Max\_Hop=3$, $p^J_u=0.15$, $p^F_u=0.85$.\\ \hline
         {NRM-VNE \cite{7976281}} &  A heuristic algorithm based on multi-dimensional resource constraints. & {Based on the MIP process.} \\ \hline
         {CDRL \cite{yao2020continuous}} & A reinforcement learning algorithm based on time series, but the embedding rules are not interpretable. & \textcolor{black}{$\mu=0.005$, $w_{cpu}=1/2$, $w_{sto}=1/2$, $w_{bw}=1$. $y_i=1$.}\\ \hline
    \end{tabular}
\end{table}
\begin{figure}
    \centering
    \includegraphics[width=0.8\linewidth]{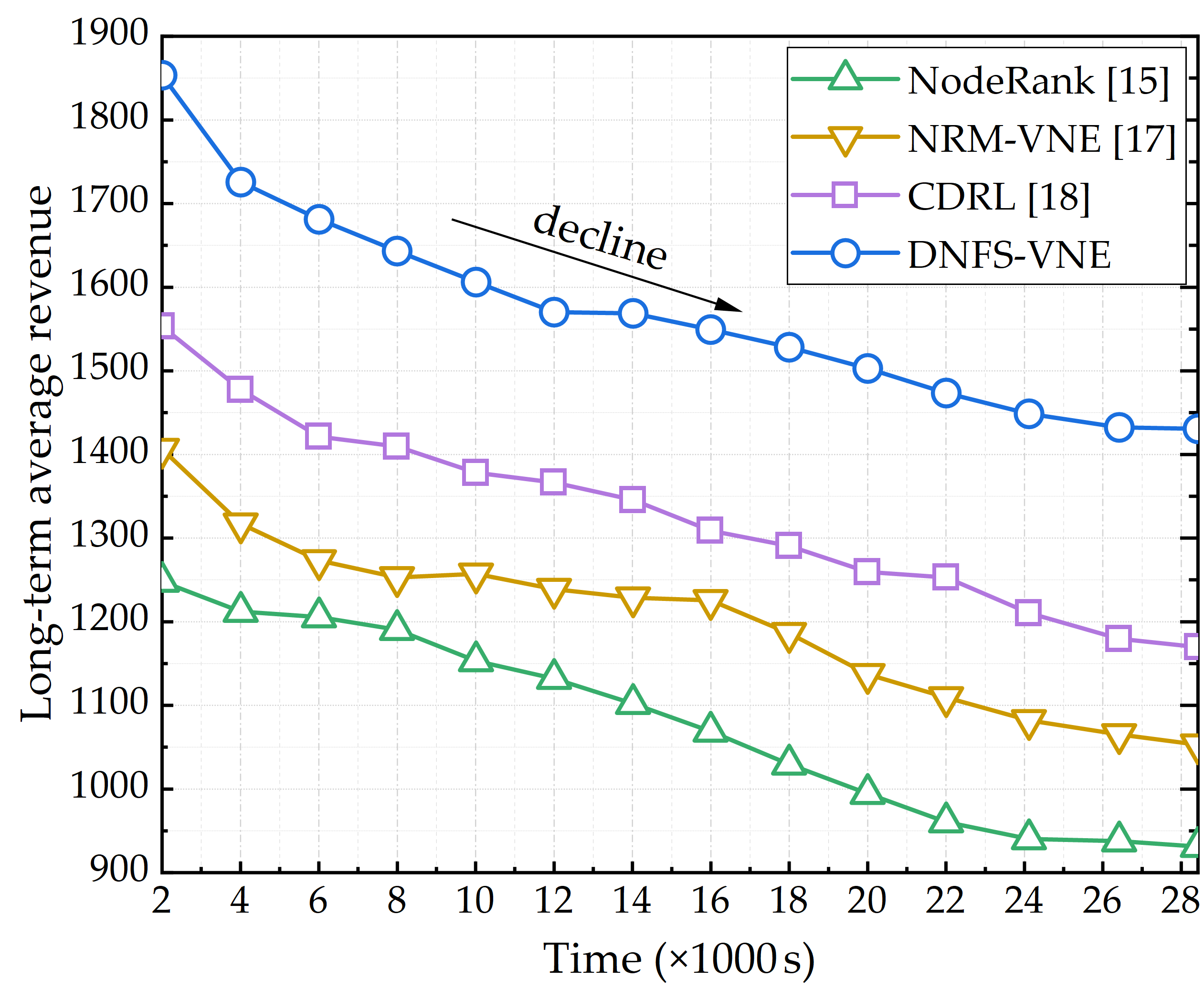}
    \vspace{-0.3cm}
    \caption{Comparison experiment on Eq.~\ref{eq5}.}
    \label{fig:7}
\end{figure}
\begin{figure}
    \centering
    \includegraphics[width=0.8\linewidth]{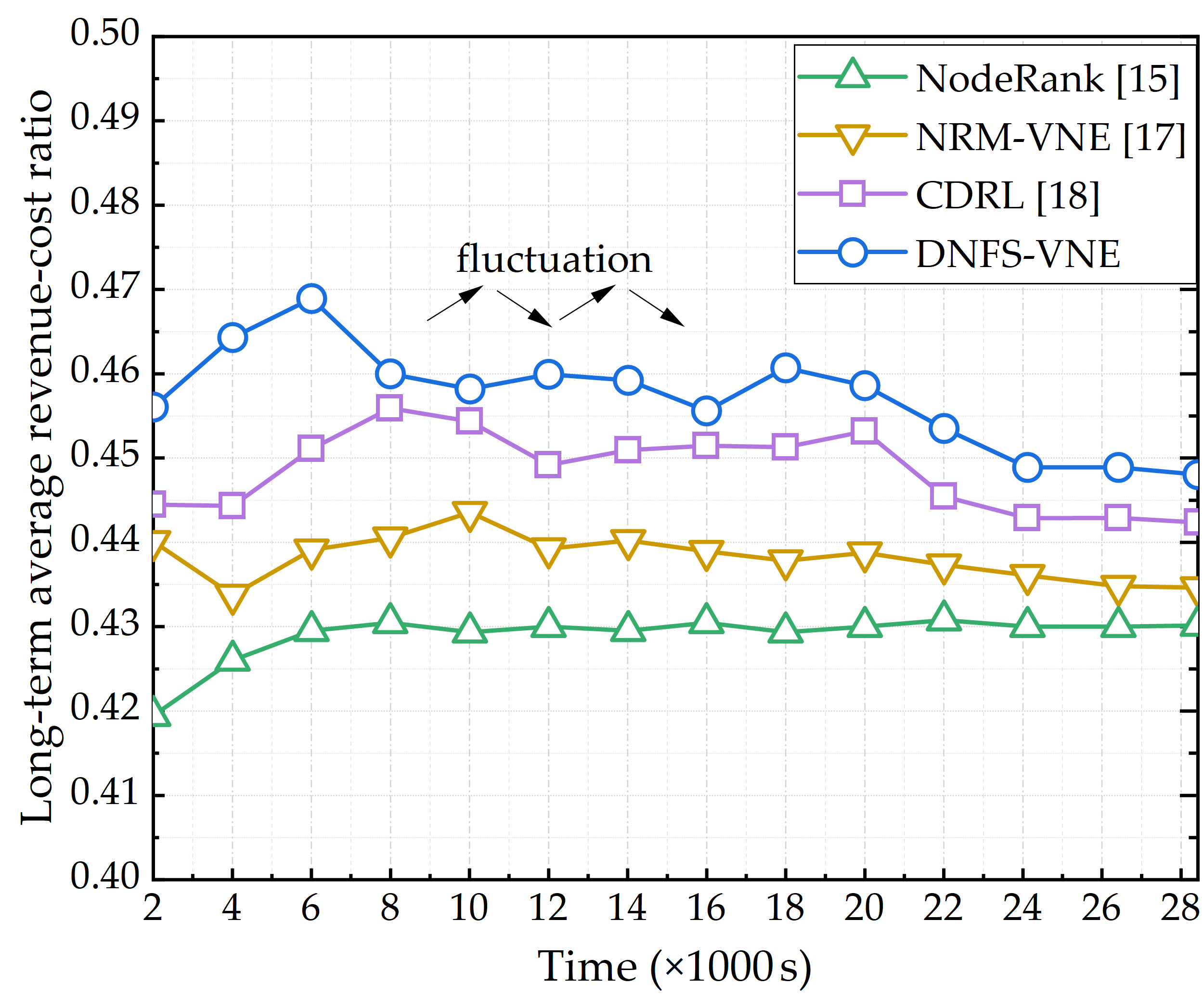}
    \vspace{-0.3cm}
    \caption{Comparison experiment on Eq.~\ref{eq6}.}
    \label{fig:8}
\end{figure}
\begin{figure}[t]
    \centering
    \includegraphics[width=0.8\linewidth]{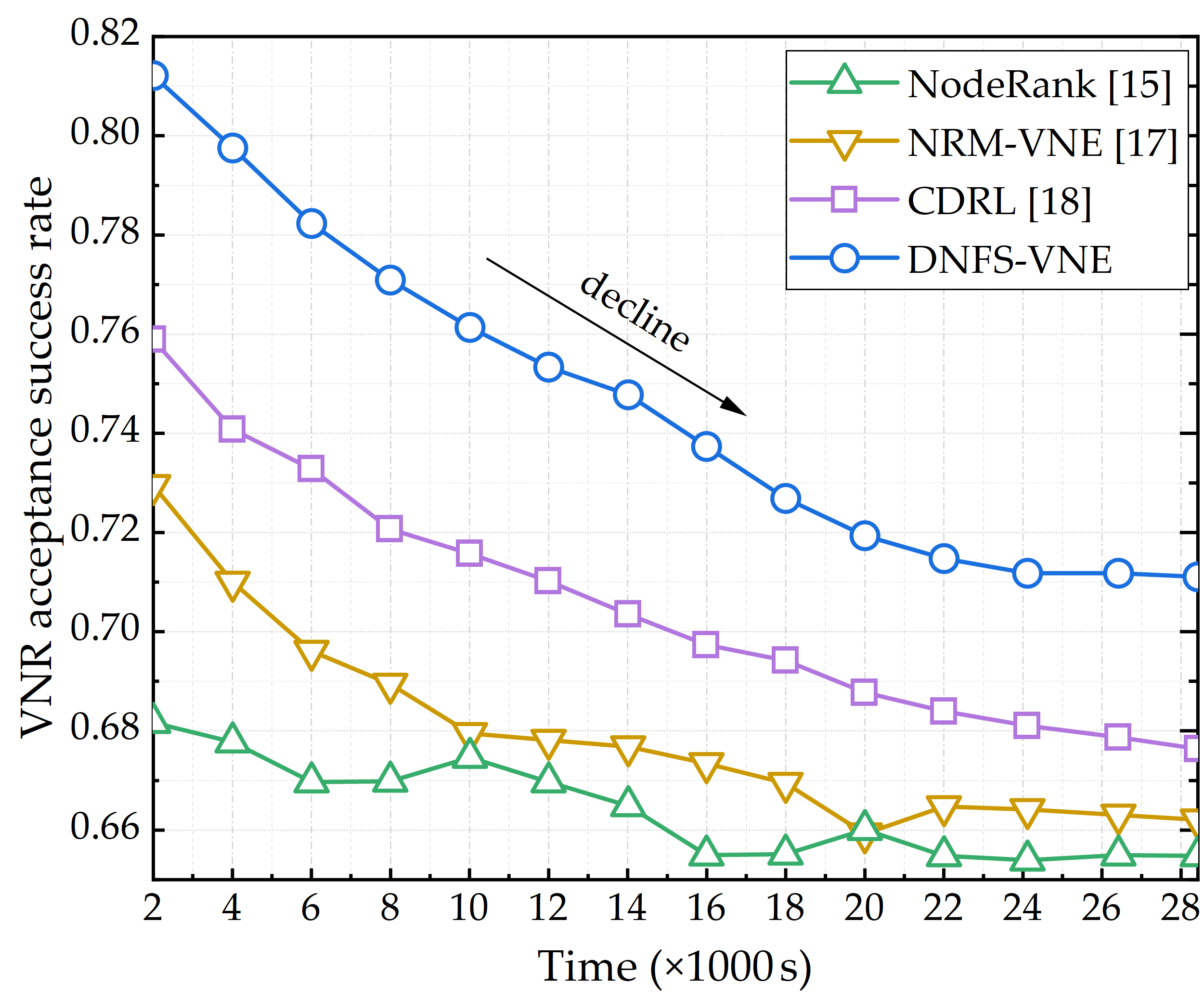}
    \vspace{-0.3cm}
    \caption{Comparison experiment on Eq.~\ref{eq7}.}
    % \vspace{-0.3cm}
    \label{fig:9}
\end{figure}

We selected the classic VNE algorithms (NodeRank~\cite{cheng2011virtual}, NRM-VNE~\cite{7976281}, CDRL~\cite{yao2020continuous}) as baselines to prove the effectiveness of DNFS-VNE. The descriptions and parameter settings of all baselines are recorded in Table~\ref{tab:para}. For rigor, all baselines are run in the same simulation environment as shown in Table~\ref{tab:conf}. Starting from the time series $22$ \text{s} of VNRs, the changes in indicators are recorded every $4,000$ \text{s}, as shown in Fig.~\ref{fig:7}, Fig.~\ref{fig:8}, and Fig.~\ref{fig:9}.

It can be found that the heuristic algorithm is generally inferior to the AI-based method. Most of the heuristic algorithms greedily allocate physical nodes and links with more resources, resulting in satisfactory performance initially but needing improvement in the long run. The CDRL algorithm introduces the RL algorithm on this basis, taking into account the interaction with the environment and comprehensively considering the VNR requirements, contributing to a significant improvement in effectiveness. In addition, the proposed DNFS-VNE fully considers the VNE implication rules and caches the identified inference patterns into weights through forward calculation and BP, thereby significantly improving the performance of each indicator.

As time progresses, the response to VNRs leads to a decrease in available resources for the substrate network. Consequently, this also reduces the number of VNRs that can be successfully embedded subsequently. Thus, the long-term average revenue and VNR acceptance success rate, as shown in Eq.~\ref{eq5} and Eq.~\ref{eq7}, will continue to decrease. For Eq.~\ref{eq6}, the long-term average revenue-cost ratio fluctuates as it is influenced by both revenue and cost. The results shown in the figures also validate these theories. In summary, this work is superior compared to other baselines as well as effective.

\section{Conclusion and Future}\label{sec5}
Based on the promising performance of interpretable DL represented by DNFS, this work proposes a DNFS-based VNE algorithm that aims to provide an interpretable NV scheme. Specifically, DNFS-VNE is a five-block architecture in which CNNs act as fuzzy implication operators to perform entailment operations and ultimately output the embedding probabilities of candidate substrate nodes. And, fuzzy rule patterns are cached into the weights during the model's forward computation and gradient back-propagation. Moreover, Mamdani-type linguistic rules are used to construct the fuzzy rule base using linguistic labels and then interpret them. Furthermore, we use the policy network based on a five-block architecture as the agent of DRL to optimize the VNE scheme through interaction with the environment and joint guidance of the reward function. Finally, the effectiveness of the algorithm is proved through experiments.

Furthermore, it is important to mention that this work is only an initial effort to explore the interpretability of the VNE algorithm through DNFS. Additionally, this work can be easily extended to incorporate more resource dimensions, such as substrate link delay, substrate node degree, \textit{etc}., to explore the influence of more attributes on the VNE process. Also, it can explore more complicated fuzzy implication designs to obtain more detailed fuzzy rules, and then develop more versions of DNFS-VNE.
\bibliographystyle{IEEEtran}
\vspace{-0.2cm}
\bibliography{reference}

\begin{thebibliography}{10}
\providecommand{\url}[1]{#1}
\csname url@samestyle\endcsname
\providecommand{\newblock}{\relax}
\providecommand{\bibinfo}[2]{#2}
\providecommand{\BIBentrySTDinterwordspacing}{\spaceskip=0pt\relax}
\providecommand{\BIBentryALTinterwordstretchfactor}{4}
\providecommand{\BIBentryALTinterwordspacing}{\spaceskip=\fontdimen2\font plus
\BIBentryALTinterwordstretchfactor\fontdimen3\font minus
  \fontdimen4\font\relax}
\providecommand{\BIBforeignlanguage}[2]{{%
\expandafter\ifx\csname l@#1\endcsname\relax
\typeout{** WARNING: IEEEtran.bst: No hyphenation pattern has been}%
\typeout{** loaded for the language `#1'. Using the pattern for}%
\typeout{** the default language instead.}%
\else
\language=\csname l@#1\endcsname
\fi
#2}}
\providecommand{\BIBdecl}{\relax}
\BIBdecl

\bibitem{10012414}
D.~Basu, S.~Kal, U.~Ghosh, and R.~Datta, ``{DRIVE: Dynamic Resource
  Introspection and VNF Embedding for 5G Using Machine Learning},'' \emph{IEEE
  Internet of Things Journal}, vol.~10, no.~21, pp. 18\,971--18\,979, January
  2023.

\bibitem{10413579}
S.~Wu, N.~Chen, G.~Wen, L.~Xu, P.~Zhang, and H.~Zhu, ``{Virtual Network
  Embedding for Task Offloading in IIoT: A DRL-Assisted Federated Learning
  Scheme},'' \emph{IEEE Transactions on Industrial Informatics}, vol.~20,
  no.~4, pp. 6814--6824, January 2024.

\bibitem{9717289}
H.~Lu and F.~Zhang, ``{Resource Fragmentation-Aware Embedding in Dynamic
  Network Virtualization Environments},'' \emph{IEEE Transactions on Network
  and Service Management}, vol.~19, no.~2, pp. 936--948, February 2022.

\bibitem{yan2020automatic}
Z.~Yan, J.~Ge, Y.~Wu, L.~Li, and T.~Li, ``{Automatic Virtual Network Embedding:
  A Deep Reinforcement Learning Approach with Graph Convolutional Networks},''
  \emph{IEEE Journal on Selected Areas in Communications}, vol.~38, no.~6, pp.
  1040--1057, April 2020.

\bibitem{10077713}
Z.~Yang, R.~Gu, H.~Li, and Y.~Ji, ``{Approximately Lossless Model
  Compression-Based Multilayer Virtual Network Embedding for Edge–Cloud
  Collaborative Services},'' \emph{IEEE Internet of Things Journal}, vol.~10,
  no.~14, pp. 13\,040--13\,055, March 2023.

\bibitem{9520818}
Y.~Li, Y.~Zuo, H.~Song, and Z.~Lv, ``{Deep Learning in Security of Internet of
  Things},'' \emph{IEEE Internet of Things Journal}, vol.~9, no.~22, pp.
  22\,133--22\,146, 2022.

\bibitem{10263775}
H.-K. Lim, I.~Ullah, J.-B. Kim, and Y.-H. Han, ``{Virtual Network Embedding
  Based on Hierarchical Cooperative Multiagent Reinforcement Learning},''
  \emph{IEEE Internet of Things Journal}, vol.~11, no.~5, pp. 8552--8568, 2024.

\bibitem{10162187}
S.~Dong and Y.~Li, ``{Adaptive Fuzzy Event-Triggered Formation Control for
  Nonholonomic Multirobot Systems With Infinite Actuator Faults and Range
  Constraints},'' \emph{IEEE Internet of Things Journal}, vol.~11, no.~1, pp.
  1361--1373, June 2024.

\bibitem{7938307}
J.-S. Lee and C.-L. Teng, ``{An Enhanced Hierarchical Clustering Approach for
  Mobile Sensor Networks Using Fuzzy Inference Systems},'' \emph{IEEE Internet
  of Things Journal}, vol.~4, no.~4, pp. 1095--1103, 2017.

\bibitem{9669061}
H.~Zhao, J.~Tang, B.~Adebisi, T.~Ohtsuki, G.~Gui, and H.~Zhu, ``{An Adaptive
  Vehicle Clustering Algorithm Based on Power Minimization in Vehicular Ad-Hoc
  Networks},'' \emph{IEEE Transactions on Vehicular Technology}, vol.~71,
  no.~3, pp. 2939--2948, January 2022.

\bibitem{8352739}
C.~Pham, L.~A.~T. Nguyen, N.~H. Tran, E.-N. Huh, and C.~S. Hong,
  ``{Phishing-Aware: A Neuro-Fuzzy Approach for Anti-Phishing on Fog
  Networks},'' \emph{IEEE Transactions on Network and Service Management},
  vol.~15, no.~3, pp. 1076--1089, April 2018.

\bibitem{talpur2023deep}
N.~Talpur, S.~J. Abdulkadir, H.~Alhussian, M.~H. Hasan, N.~Aziz, and A.~Bamhdi,
  ``{Deep Neuro-Fuzzy System Application Trends, Challenges, and Future
  Perspectives: A Systematic Survey},'' \emph{Artificial intelligence review},
  vol.~56, no.~2, pp. 865--913, April 2023.

\bibitem{drone2023}
N.~Chen, S.~Shen, Y.~Duan, S.~Huang, W.~Zhang, and L.~Tan, ``{Non-Euclidean
  Graph-Convolution Virtual Network Embedding for Space–Air–Ground
  Integrated Networks},'' \emph{Drones}, vol.~7, no.~3, p. 165, February 2023.

\bibitem{9766416}
L.~Song, X.~Hu, G.~Zhang, P.~Spachos, K.~N. Plataniotis, and H.~Wu,
  ``{Networking Systems of AI: On the Convergence of Computing and
  Communications},'' \emph{IEEE Internet of Things Journal}, vol.~9, no.~20,
  pp. 20\,352--20\,381, May 2022.

\bibitem{cheng2011virtual}
X.~Cheng, S.~Su, Z.~Zhang, H.~Wang, F.~Yang, Y.~Luo, and J.~Wang, ``{Virtual
  Network Embedding through Topology-Aware Node Ranking},'' \emph{ACM SIGCOMM
  Computer Communication Review}, vol.~41, no.~2, pp. 38--47, April 2011.

\bibitem{5061987}
N.~M. M.~K. Chowdhury, M.~R. Rahman, and R.~Boutaba, ``{Virtual Network
  Embedding with Coordinated Node and Link Mapping},'' in \emph{IEEE INFOCOM
  2009}, June 2009, pp. 783--791.

\bibitem{7976281}
P.~Zhang, H.~Yao, and Y.~Liu, ``{Virtual Network Embedding Based on Computing,
  Network, and Storage Resource Constraints},'' \emph{IEEE Internet of Things
  Journal}, vol.~5, no.~5, pp. 3298--3304, July 2018.

\bibitem{yao2020continuous}
H.~Yao, S.~Ma, J.~Wang, P.~Zhang, C.~Jiang, and S.~Guo, ``{A
  Continuous-Decision Virtual Network Embedding Scheme Relying on Reinforcement
  Learning},'' \emph{IEEE Transactions on Network and Service Management},
  vol.~17, no.~2, pp. 864--875, February 2020.

\bibitem{zhang2022blockchain}
S.~Zhang, Z.~Wang, Z.~Zhou, Y.~Wang, H.~Zhang, G.~Zhang, H.~Ding, S.~Mumtaz,
  and M.~Guizani, ``{Blockchain and Federated Deep Reinforcement Learning based
  Secure Cloud-Edge-end Collaboration in Power IoT},'' \emph{IEEE Wireless
  Communications}, vol.~29, no.~2, pp. 84--91, April 2022.

\bibitem{8845171}
M.~Dolati, S.~B. Hassanpour, M.~Ghaderi, and A.~Khonsari, ``{DeepViNE: Virtual
  Network Embedding with Deep Reinforcement Learning},'' in \emph{IEEE INFOCOM
  2019 - IEEE Conference on Computer Communications Workshops (INFOCOM
  WKSHPS)}, September 2019, pp. 879--885.

\bibitem{CHEN2022109931}
N.~Chen, P.~Zhang, N.~Kumar, C.-H. Hsu, L.~Abualigah, and H.~Zhu, ``{Spectral
  Graph Theory-Based Virtual Network Embedding for Vehicular Fog Computing: A
  Deep Reinforcement Learning Architecture},'' \emph{Knowledge-Based Systems},
  vol. 257, p. 109931, Decemeber 2022.

\bibitem{10132867}
P.~Zhang, N.~Chen, S.~Li, K.-K.~R. Choo, C.~Jiang, and S.~Wu, ``{Multi-Domain
  Virtual Network Embedding Algorithm Based on Horizontal Federated
  Learning},'' \emph{IEEE Transactions on Information Forensics and Security},
  vol.~18, pp. 3363--3375, May 2023.

\bibitem{fischer2013virtual}
A.~Fischer, J.~F. Botero, M.~T. Beck, H.~De~Meer, and X.~Hesselbach, ``{Virtual
  Network Embedding: A Survey},'' \emph{IEEE Communications Surveys \&
  Tutorials}, vol.~15, no.~4, pp. 1888--1906, February 2013.

\bibitem{10040224}
W.~Fan, F.~Xiao, M.~Lv, L.~Han, J.~Wang, and X.~He, ``{Node Essentiality
  Assessment and Distributed Collaborative Virtual Network Embedding in
  Datacenters},'' \emph{IEEE Transactions on Parallel and Distributed Systems},
  vol.~34, no.~4, pp. 1265--1280, February 2023.

\bibitem{10144650}
J.~Zhu, W.~Zhao, H.~Yang, and F.~Nie, ``{Joint Learning of Anchor Graph-Based
  Fuzzy Spectral Embedding and Fuzzy K-Means},'' \emph{IEEE Transactions on
  Fuzzy Systems}, vol.~31, no.~11, pp. 4097--4108, June 2023.

\bibitem{pang2022fuzzy}
B.~Pang, ``{Fuzzy Convexities Via Overlap Functions},'' \emph{IEEE Transactions
  on Fuzzy Systems}, vol.~31, no.~4, pp. 1071--1082, July 2022.

\bibitem{cao2017novel}
H.~Cao, Y.~Zhu, G.~Zheng, and L.~Yang, ``{A Novel Optimal Mapping Algorithm
  with Less Computational Complexity for Virtual Network Embedding},''
  \emph{IEEE Transactions on Network and Service Management}, vol.~15, no.~1,
  pp. 356--371, November 2017.

\bibitem{9894092}
S.~Ma, H.~Yao, T.~Mai, J.~Yang, W.~He, K.~Xue, and M.~Guizani, ``{Graph
  Convolutional Network Aided Virtual Network Embedding for Internet of
  Thing},'' \emph{IEEE Transactions on Network Science and Engineering},
  vol.~10, no.~1, pp. 265--274, September 2023.

\bibitem{likas2003global}
A.~Likas, N.~Vlassis, and J.~J. Verbeek, ``{The Global K-Means Clustering
  Algorithm},'' \emph{Pattern recognition}, vol.~36, no.~2, pp. 451--461,
  February 2003.

\bibitem{iancu2012mamdani}
I.~Iancu, ``{A Mamdani Type Fuzzy Logic Controller},'' \emph{Fuzzy
  logic-controls, concepts, theories and applications}, vol.~15, no.~2, pp.
  325--350, March 2012.

\bibitem{10132506}
R.~Zhu, G.~Li, Y.~Zhang, Z.~Fang, and J.~Wang, ``{Load-Balanced Virtual Network
  Embedding Based on Deep Reinforcement Learning for 6G Regional Satellite
  Networks},'' \emph{IEEE Transactions on Vehicular Technology}, vol.~72,
  no.~11, pp. 14\,631--14\,644, May 2023.

\end{thebibliography}
\end{document}